\documentclass[hyperpdf,aps,prd,superscriptaddress,nofootinbib,preprint,preprintnumbers,floatfix,tightenlines]{revtex4-1}

\pdfoutput=1   

\usepackage{amsfonts}
\usepackage{amsmath}
\usepackage{amssymb}
\usepackage{bm}
\usepackage{graphicx}
\usepackage[pdftex]{color}
\usepackage[sort&compress]{natbib}
\usepackage[colorlinks=true,linkcolor=blue,filecolor=blue,urlcolor=blue,citecolor=blue,pdftex,plainpages=false]{hyperref}
\usepackage{sidecap}
\usepackage{wrapfig}

\usepackage[utf8]{inputenc}

\begin{document}

\title{Hadrons and Nuclei}
\author{William Detmold (editor)}
\affiliation{Massachusetts Institute of Technology}
\author{Robert G. Edwards (editor)}
\affiliation{Jefferson Lab}
\author{Jozef J. Dudek}
\affiliation{Jefferson Lab}
\affiliation{College of William and Mary}
\author{Michael Engelhardt}
\affiliation{New Mexico State University}
\author{Huey-Wen Lin}
\affiliation{Michigan State University}
\author{Stefan Meinel}
\affiliation{University of Arizona}
\affiliation{RIKEN BNL Research Center}
\author{Kostas Orginos}
\affiliation{Jefferson Lab}
\affiliation{College of William and Mary}
\author{Phiala Shanahan}
\affiliation{Massachusetts Institute of Technology}

\begin{abstract}

This document is one of a series of whitepapers from the USQCD collaboration.  Here, we discuss opportunities for lattice QCD calculations related to the structure and spectroscopy of hadrons and nuclei. An overview of recent lattice calculations of the structure of the proton and other hadrons is presented along with prospects for future extensions. Progress and prospects of hadronic spectroscopy and the study of resonances in the light, strange and heavy quark sectors is summarized. Finally, recent advances in the study of light nuclei from lattice QCD are addressed, and the scope of future investigations that are currently envisioned is outlined.

\end{abstract}

\collaboration{USQCD Collaboration}

\maketitle

\tableofcontents
\newpage 

\section*{Executive summary}
In 2018, the USQCD collaboration's Executive Committee organized several subcommittees to recognize future opportunities and formulate possible goals for lattice field theory calculations in several physics areas. The conclusion of these studies, along with community input, are presented in seven whitepapers \cite{Bazavov:2018qcd,Brower:2018qcd,Davoudi:2018qcd,Detmold:2018qcd,Joo:2018qcd,Kronfeld:2018qcd,Lehner:2018qcd}. Here, we discuss opportunities for lattice QCD calculations related to the structure and spectroscopy of hadrons and nuclei. 

Nuclear Physics is a diverse field with linkages to many areas of
research and experimentation, including the structure of  hadrons
and the properties of the nuclei composed of protons and neutrons.
There are existing and new generations of experiments within the US,
and also worldwide, dedicated to explaining these properties as laid
out in the 2015 NSAC Long Range Plan for Nuclear Physics.  The
RHIC-spin program (BNL), the recent
12GeV upgrade of Jefferson Lab (JLab) and the planned Electron Ion
Collider (EIC), amongst others, will peer into the internal structure of hadrons and
look for the possible existence of exotic states of matter. The
Facility for Rare Isotopes program (FRIB) will clarify how subatomic
matter organizes itself and how nuclei emerge.

This impressive level of experimentation has resulted in
numerous discoveries that have led to the development of the
fundamental theory that describes the strong interactions -- Quantum
Chromodynamics (QCD). This theory, when combined with the electroweak
interactions, underlies all of nuclear physics, from the spectrum and
structure of hadrons to the most complex nuclear reactions. However,
many aspects of nuclear physics are dictated by the regime of QCD in
which its defining feature--asymptotic freedom--is concealed by
confinement and by the complicated structure of the quantum
vacuum. The numerical technique of {\it Lattice QCD} is the only known
way to perform {\it ab initio} QCD calculations of strong interaction
quantities in this regime. The ability to compute the properties of
matter, with quantifiable uncertainties, is necessary to establish
a bridge between theory and experiments, and vital to progress in the
field.  

Lattice QCD (LQCD) is a technique in which space and time are
discretized and strong interaction quantities are calculated by
large-scale numerical Monte-Carlo integration, and in which
approximation effects can be systematically removed. The LQCD community
has been at the forefront of innovation in, and utilization of, high performance computing  for
decades, and the ambitious plans put forth in this white paper will
require still larger computing capabilities. To this end, the SciDAC
programs have been essential to achieving high performance on new hardware
architectures, and LQCD calculations have led the development and
adoption of new computing paradigms, including the use of graphical
processing units. Local computing resources under the USQCD Initiative
have also been essential to effectively using the
leadership facilities. In the near future, Exascale computing resources
will be required, and the software development efforts under the
Exascale Computing Project and SciDAC-4 program are paving the way for new calculations
beyond those currently possible.

This white paper provides a roadmap for on-going and future science
programs that will have a profound impact on our understanding of hadrons and nuclei.

\section{Introduction}
\label{sec:intro}
Hadrons and nuclei make up the bulk of everyday matter and are the objects that are detected in experiments at accelerators and colliders. Yet hadrons and nuclei are complex composite objects that emerge from the underlying strong interactions between quarks and gluons, the fundamental degrees of freedom of Quantum Chromodynamcis (QCD). Describing this compositeness is challenging as the couplings between quarks and gluons become large at the relevant energy scales, and the perturbative approach that works well for QED and for QCD at high energy breaks down.
In this USQCD collaboration whitepaper, we discuss the application of the numerical techniques of lattice QCD (LQCD) to calculations of the non-perturbative properties of hadrons and nuclei. We summarize the  recent accomplishments of LQCD (and USQCD in particular) in this domain and discuss future goals and opportunities in the context of current and future experiments. There  are numerous synergies between the topics discussed here and those discussed in the six companion USQCD whitepapers
\cite{Bazavov:2018qcd,Brower:2018qcd,Davoudi:2018qcd,Joo:2018qcd,Kronfeld:2018qcd,Lehner:2018qcd}.
which are highlighted in the following.

\section{Hadron Structure}
\label{sec:hadronstructure}

The study of the structure of the proton and other hadrons is a central pursuit in nuclear physics. Since the 1950s, probing this structure has revealed new aspects of nature and ultimately led to the development of the Standard Model. The pioneering experiments of Hofstadter revealed the charge distribution of the proton and nuclei, while the deep-inelastic scatting (DIS) experiments by Kendall, Taylor and Friedman at SLAC and those that followed led to the development of QCD and have mapped out  the distributions of fundamental partonic (quark and gluon) degrees of freedom in the proton. At present, these studies are being complemented by new generations of experiments such as those at the 12 GeV upgrade of Jefferson Lab and a potential Electron-Ion Collider (EIC) that seek to map out the three-dimensional structure of the proton by determining generalized parton distributions and transverse-momentum dependent parton distributions. 

Lattice QCD studies of hadron structure were pioneered in the 1980s and have matured significantly. They are now of a maturity where  rigorous connection to experiment is possible and the prospect of future calculations that will support and complement modern experimental investigations is exciting.

\subsection{Charges, radii,  electroweak form factors and polarizabilities}

The simplest aspects of hadron structure that are probed in electroweak interactions are the various static ``charges'' and moments corresponding to the coupling of bilinear quark currents to the hadron. Generalising to currents involving momentum transfer leads to the  electroweak form factors, with the small momentum behaviour characterized by the electromagnetic and weak radii that correspond to the slopes of the appropriate form factors at zero momentum transfer. These quantities can be determined from LQCD by calculating ratios of three-point and two point correlations functions built from hadronic interpolating operators and quark current operators. This is by now a well-developed approach with various groups around the world presenting results that are close to controlling all systematic uncertainties. Over the next few years, these systematic uncertainties will be further reduced  and  precision will be increased by performing high statistics calculations with additional lattice spacings and volumes. In addition, the full flavor-dependence of the moments, radii and form factors will be determined. 

There are a number of particularly important cases in this class of LQCD calculations. 
\begin{itemize}
	\item The axial charge of the nucleon is a benchmark quantity that is known very precisely from experiment, $g_A=1.2723(23)$ \cite{Patrignani:2016xqp}. LQCD calculations are also becoming more precise, with uncertainties at the few percent level  \cite{Bhattacharya:2016zcn,Yoon:2016jzj,Chang:2018uxx,Gupta:2018qil}. With the significant increase in precision that will occur in the next few years, it is possible that this will  become a quantity that tests the Standard Model; this is particularly relevant in the context of current anomalies in different measurements of the neutron lifetime. Recent calculations of this quantity by USQCD collaboration members are becoming increasingly precise.
	
	\item The scalar charge of the nucleon dictates the sensitivity of searches for important classes of dark matter candidates at direct-detection experiments. Along with the tensor charge \cite{Gupta:2018lvp}, the scalar charge \cite{Shanahan:2016pla}  is also relevant theory input to other searches for physics beyond the Standard Model. These quantities are discussed further in the companion white paper on Fundamental Symmetries \cite{Davoudi:2018qcd}.
	
	\item The proton charge radius is  of significant phenomenological interest as there are very significant discrepancies in its extraction from muonic hydrogen spectroscopy  and from electronic hydrogen spectroscopy and electron-proton scattering. Existing LQCD calculations of the isovector charge radius of the proton \cite{Capitani:2015sba, Hasan:2017wwt,Alexandrou:2017ypw,Ishikawa:2018rew,Detmold:2018ptb,Alexandrou:2018sjm}  have $\sim10$\% precision, assigning conservative estimates of the systematic uncertainties that are not well-quantified. A  percent-level  LQCD calculation of the isovector radius combined with existing precise measurements of the charge radius of the neutron are sufficient to determine the proton charge radius at a level where LQCD calculations will  have impact on the discrepancy. 
	
	\item The axial current form factors of the nucleon and nuclei are relevant for neutrino physics as discussed extensively in the  accompanying whitepaper on lattice QCD for neutrino physics \cite{Kronfeld:2018qcd}.
\end{itemize} 
These quantities are relatively simple to calculate  and have been analyzed by the community in Ref.~\cite{Lin:2017snn} and  are being included in the upcoming 2018 Flavor Lattice Averaging Group (FLAG {\tt http://flag.unibe.ch})
review of LQCD calculations.

Second order responses to EM fields are quantified by the electric and magnetic (and higher-order spin) polarizabilities of hadrons. LQCD calculations of polarizabilities have used spectroscopy in fixed external fields~\cite{Savage:2016kon,Shanahan:2017bgi,Tiburzi:2017iux} or direct measurement of hadronic four point functions corresponding to two current insertions~\cite{Engelhardt:2007ub}. Being somewhat complicated observables, polarizability calculations with close to physical quark masses and with explicit control of all systematic uncertainties are lacking but will be possible with the levels of resources available in the next five years.

\subsection{Parton Distribution Functions}

The DIS experiments begun at SLAC in the late 1960s, led the way to the observation of asymptotic freedom and the development of QCD as a non-Abelian gauge theory. Efforts to better determine the partonic structure seen inside the proton have continued ever since. The parton distributions functions (PDFs), which quantify the densities of quarks and gluons in a hadron as a function of the longitudinal momentum fraction, $x$, are important inputs for experiments at hadron colliders such as the LHC and must be better constrained to fully exploit these experimental programs.
They  are defined by matrix elements in a hadron state of bi-local operators separated along the light-cone and are intrinsically difficult to access from LQCD calculations in Euclidean space ~\cite{Collins:1981uw,Curci:1980uw,Baulieu:1979mr,Collins:1989gx} .

\subsubsection{Moments of Parton Distribution Functions}

The most well-established computations that address the partonic structure of hadrons are based on calculations of matrix elements of the local twist-two operators that arise in the light-cone operator product expansion (OPE) of DIS and related processes. These matrix elements determine the Mellin moments of the underlying parton distributions; with  a sufficient number of moments the PDF can be reconstructed with controlled uncertainties. However, the reduced symmetries of the spacetime lattice used in LQCD calculations (typically, the hypercubic group H(4)) compared to the Lorentz group means that the OPE is complicated by divergent mixing between operators.
Lattice operators corresponding to the lowest few moments of the unpolarized, polarized and transversity distributions can be chosen such that this mixing is absent at the expense of having nonzero matrix elements only in states of nonzero three-momentum. LQCD calculations of these matrix elements have been undertaken since the first calculations of Martinelli and Sachrajda \cite{Dawson:1997ic} in the 1980s. A recent summary of the calculations of PDF moments is given in \cite{Lin:2017snn}.
	
To go beyond the lowest moments, operators involving more complicated finite difference discretizations of the derivative operators can be constructed following ideas developed in Ref.~\cite{Davoudi:2012ya} for  three-dimensional discretizations of interpolating operators of fixed angular momentum. By using multiple copies of given irreducible H(4) representations (irreps), better approximations to operators transforming irreducibly under SO(4) symmetries can be constructed. This approach is being actively investigated at present \cite{IDavoudiLattice2018} and offers the possibility of calculations of sufficient numbers of moments that a parameterization of the underlying PDF can be constrained.

\subsubsection{Quasi-distributions and pseudo-distributions}

\begin{figure}[h!]
	\centering
	\includegraphics[width=0.45\columnwidth]{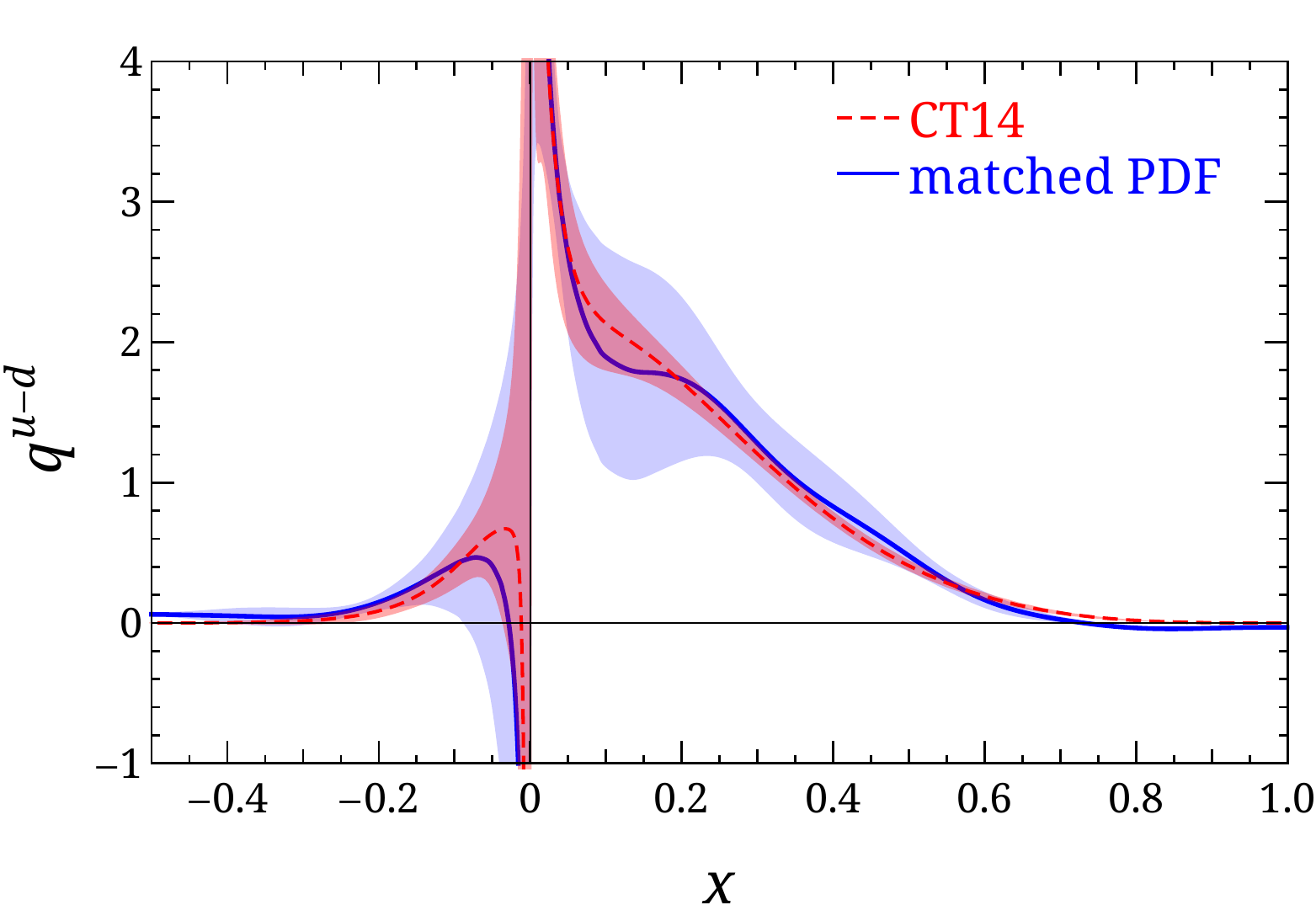}\hspace{1cm}
	\includegraphics[width=0.45\columnwidth]{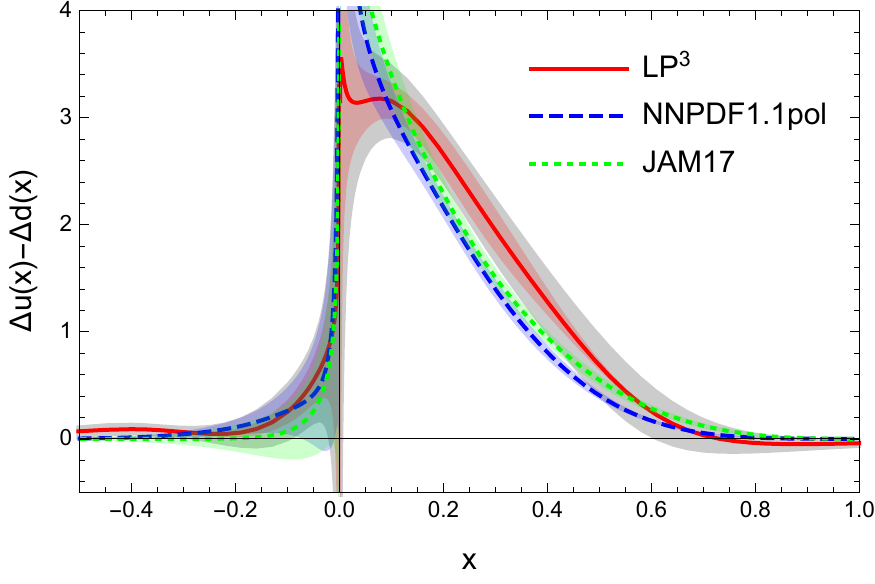}
	\caption{Left: Unpolarized isovector nucleon PDF with comparison to the CTEQ parameterization~\cite{Chen:2018xof}.
		Right: Polarized isovector nucleon PDF with comparison to the NNPDF parameterization~\cite{Lin:2018qky}.}
	\label{fig_quasipdf}
\end{figure}

The formulation of lattice QCD in Euclidean space severely restricts
lattice calculations of partonic structure.  
The analytic continuation of the matrix elements that define the PDFs to Euclidean space is highly non-trivial due to the fact that these matrix elements are not local in time. Recently, new ideas,  known as the "Large Momentum Effective Field Theory" (LaMET),  have
been proposed that aim to circumvent this problem~\cite{Ji:2013dva,Ji:2014gla}.
In this approach, one computes a time local version of the matrix element that defines the PDF in Euclidean space
where the external states have a suitably large momentum  and the
bi-local quark insertion is separated by some spatial distance.
With these choices, the quasi-PDF is defined by the Fourier transform over the spatial extent of the equal-time matrix element of a spatially directed Wilson-line between quark fields, at some lattice scale. To relate this lattice quasi-PDF to the desired Minkowski light-cone PDF, matching conditions are implemented within the LaMET~\cite{Ji:2013dva,Ji:2014gla} scheme after either perturbative or non-perturbative~\cite{Martinelli:1994ty} renromalization. Power corrections that break the matching  procedure from  higher-twist effects are suppressed at large nucleon momentum. This approach has been recently used  for quasi-PDFs  in Refs.~\cite{Alexandrou:2017huk,Chen:2017mzz,Green:2017xeu,Chen:2018xof,Lin:2018qky}. A recent determination of the isovector unpolarized and polarized PDFs of the  nucleon is shown in Figure~\ref{fig_quasipdf}.

An alternative approach, named the pseudo-PDF, considers the ratio of the equal time matrix element of the Wilson line between quarks with the rest-frame density matrix element. The equal time matrix element is parameterized in terms of the product of the spatial momentum with the spatial separation forming a Lorentz invariant called the Ioffe time~\cite{Ioffe:1969kf,Braun:1994jq}, and the ratio corresponds to the Ioffe time distribution~\cite{Radyushkin:2016hsy,Radyushkin:2017cyf}.
This ratio is free of UV divergences and requires no renormalization. The key distinction between the quasi and pseudo PDF approaches is that in the latter the Fourier transform over all spatial separations is not, in practice, needed. Indeed, recent work has shown that there can be large finite-volume effects within the spatial integration~\cite{Briceno:2018lfj}. An observation from initial lattice calculations using the pseudo-PDF approach~\cite{Karpie:2018zaz,Karpie:2018zaz}  is that Ioffe-time distributions exhibit factorization down to small distances in the spatial separation, where the small distance behavior of the pseudo-PDF  satisfies a perturbative  evolution equation. Thus, rather than computing the entire PDF as a function of Bjorken-$x$, the PDF is parameterized as a function of $x$, similar to approaches taken in phenomenological studies~\cite{Ball:2017nwa,Accardi:2016qay}. Larger lattice sizes with smaller lattice spacing will allow for better probes of the perturbative evolution scale, and better constraint of the small-$x$ region.

\subsubsection{Good lattice cross sections}
Analogous to extracting PDFs from QCD global fits of high energy scattering data, PDFs can also be extracted from analyzing ``data'' generated by LQCD calculation of good {\it lattice cross sections} \cite{Ma:2014jla,Ma:2014jga}. A {\it lattice cross section} is defined in Refs.~\cite{ Ma:2014jla,Ma:2014jga} as a single-hadron matrix element of a time-ordered, renormalized nonlocal operator ${\cal O}_n(z)$: ${\sigma}_{n}(\nu,z^2,p^2)=\langle p| {T}\{{\cal O}_n({z})\}|p\rangle$ with four-vector momentum, $p$, antiquark quark-pair separation $z$, and $\nu\equiv p\cdot z$. The values of $p$ and $z$, and the choice of ${\cal O}_n$, determine the dynamical features of the lattice cross section. A useful lattice cross section should have the following three key properties: (1) calculable in LQCD in Euclidean time, (2) has a well-defined continuum limit as the lattice spacing $a\to 0$, and (3) has the same factorizable logarithmic collinear divergences as that of PDFs, which connects the good lattice cross sections to PDFs, just as high energy hadronic cross sections are related to PDFs in terms of QCD factorization.  

A class of { good} lattice cross sections was constructed in terms of a correlation of the product of two {renormalizable} currents (see also the following subsection).  There are many possible choices for the current, such as a vector quark current, or a tensor gluonic current~\cite{Ma:2017pxb}.  Different combinations of the two currents help enhance the lattice cross sections' flavor dependence.  If spatial separation between the currents is sufficiently small, the lattice cross section constructed from two renormalizable currents can be factorized into PDFs and perturbative hard kernels \cite{Ma:2017pxb},
and the PDFs can be extracted from global fits of lattice-QCD generated data for various lattice cross sections $\sigma_{n}(\nu,z^2,p^2)$ with corresponding perturbatively calculated coefficients.

The quasi-PDFs and pseudo-PDFs introduced above are derived from choosing a single anti-quark quark pair separated in space by a Wilson line.
With two space separated currents, modulo $O(\alpha_s)$ and higher twist corrections, one finds that a quasi-quark distribution is obtained when a cross-section is computed for fixed momenta, while the pseudo-quark distribution is object if the cross-section is computed with fixed spatial separation of the currents.  
That is, these two approaches for extracting PDFs are equivalent if matching coefficients are calculated at the lowest order in $\alpha_s$ neglecting all power corrections, but different if contributions from either higher order in $\alpha_s$ or higher powers in $z^2$ are considered.

\subsubsection{Hadronic tensor methods}

A variety of other approaches are also being investigated to access hadronic structure based on computations of the hadronic tensor \cite{Liu:1993cv,Aglietti:1998mz,Detmold:2005gg,Liu:2016djw}. In the first of these approaches \cite{Liu:1993cv,Liu:2016djw}, partonic physics is accessed through a discrete Laplace transform of the Euclidean hadronic tensor. Various implementations of the challenging inverse problem that is involved have been investigated in \cite{Liang:2017mye}. In the second approach, a fictitious heavy quark field is introduced and the corresponding hadronic tensor involving heavy-light currents and resulting lattice correlation functions are matched on to the relevant OPE to extract the moments of regular parton distributions. This approach requires very fine discretization scales, but first investigations are now beginning \cite{Detmold:2018kwu}. 
An additional approach based on transforms of the hadronic tensor is being pursued in Refs.~\cite{Chambers:2017dov}.

\subsection{Generalized Parton Distribution Functions}

Generalized parton distributions (GPDs) \cite{Ji:2001wha,Radyushkin:1997ki,Diehl:2003ny,Belitsky:2005qn} provide further insight into the quark and gluon structure of hadrons, combining parton dependence on longitudinal and transverse position (when viewed in their impact-parameter space formulation \cite{Burkardt:2000za}). GPDs are defined as  off-forward matrix elements of the same operators that define parton distributions.   Information about GPDs is accessible from deeply virtual Compton scattering and deeply-virtual vector meson production in particular. Basic aspects of these distributions have been investigated at JLab, COMPASS and HERMES and a significant fraction of the experimental program at the 12GeV upgrade of Jefferson Lab is focused on revealing more  information about  GPDs. Lattice calculations have focused on the generalized form factors (GFFs) that parametrize off-forward matrix elements of local twist-two operators and correspond to moments of GPDs \cite{Hagler:2003jd,Hagler:2007xi,Hagler:2009mb}. 

For the unpolarized case
GPDs also encode the spin decomposition of the proton through Ji's sum rule \cite{Ji:1996ek} that separates quark spin, orbital angular momentum and the total gluon angular momentum. 
A further decomposition, known as the Jaffe-Manohar decomposition \cite{Jaffe:1989jz}, is valid in light-cone gauge. These decompositions of the proton spin has recently been investigated in Refs.~\cite{Yang:2016plb,Alexandrou:2017oeh}.
The $n=2$ GFFs  parameterize the matrix elements of the energy momentum tensor. As well as determining the momentum distribution and spin, they 
also define the pressure and shear distributions in the hadron~\cite{Polyakov:2018zvc}.

The first calculations of the quark GFFs for $n=2,3$ were performed by USQCD collaboration members in Refs.~\cite{Hagler:2003jd} with many subsequent improvements (see Ref.~\cite{Hagler:2009mb} for a review). The isovector combination of the unpolarized and polarized GFFs have been studied at quark masses corresponding to $m_\pi>200$ MeV, but not yet at the physical point \cite{Syritsyn:2011vk,Bali:2013dpa,Hagler:2007xi,Bratt:2010jn,Sternbeck:2012rw,Brommel:2007sb,Gockeler:2003jfa,Alexandrou:2011nr}. In most calculations of the isoscalar GFFs, the disconnected contractions have been omitted with the notable exception of Ref.~\cite{Deka:2013zha}, and mixing of the quark operators with gluon operators has been ignored given the small size of perturbative mixing coefficients \cite{Alexandrou:2016ekb}.  Calculations in the next few years will address these quantities with high fidelity, controlling all systematic uncertainties.

\subsection{Transverse momentum-dependent parton distributions}
\label{TMDs}

Transverse momentum-dependent parton distributions (TMDs) \cite{Boer:2011fh}
constitute one of the pillars on which the three-dimensional tomography of
hadrons rests. Together with the three-dimensional spatial information
derived from GPDs, they permit a
comprehensive reconstruction of hadron substructure and thus
encode 
orbital angular momentum contributions to nucleon spin, and
spin-orbit correlations in hadrons. Through the selection of
particular parton spin and transverse momentum components, a
variety of TMDs can be probed, including naively time-reversal odd
(T-odd) quantities such as the Sivers and Boer-Mulders functions.
These latter TMDs exist by virtue of initial or final state interactions
in corresponding physical processes, introducing a preferred chronology
in the description of the process.

In view of the fundamental importance of TMDs and the rich spectrum of
effects that can be probed, TMDs have been, and continue to be the
target of a variety of experimental efforts. Deep-inelastic scattering
experiments performed by COMPASS \cite{Alekseev:2008aa,Adolph:2014fjw},
HERMES \cite{Airapetian:2009ae,Airapetian:2013bim} and Jefferson Lab
\cite{Qian:2011py,Avakian:2010ae} have yielded TMD data including evidence
for the T-odd Sivers effect. Complementary Drell-Yan experiments at
COMPASS \cite{Gautheron:2010wva} and Fermilab \cite{Brown:2014sea} are
envisaged, which could, in particular, test the sign change between
the SIDIS and DY processes. Related transverse single-spin
asymmetries have been measured at RHIC in polarized proton-proton
collisions \cite{Adare:2013ekj,Adamczyk:2012xd}. Further experimental
efforts at RHIC are projected to provide insight into strong QCD evolution
effects expected for the Sivers TMD \cite{Echevarria:2014vda}.
TMDs furthermore constitute a central focus of the proposed Electron-Ion
Collider facility \cite{Accardi:2012qut}.

\begin{figure}[h!]
	\centering
	\includegraphics[width=0.474\columnwidth]{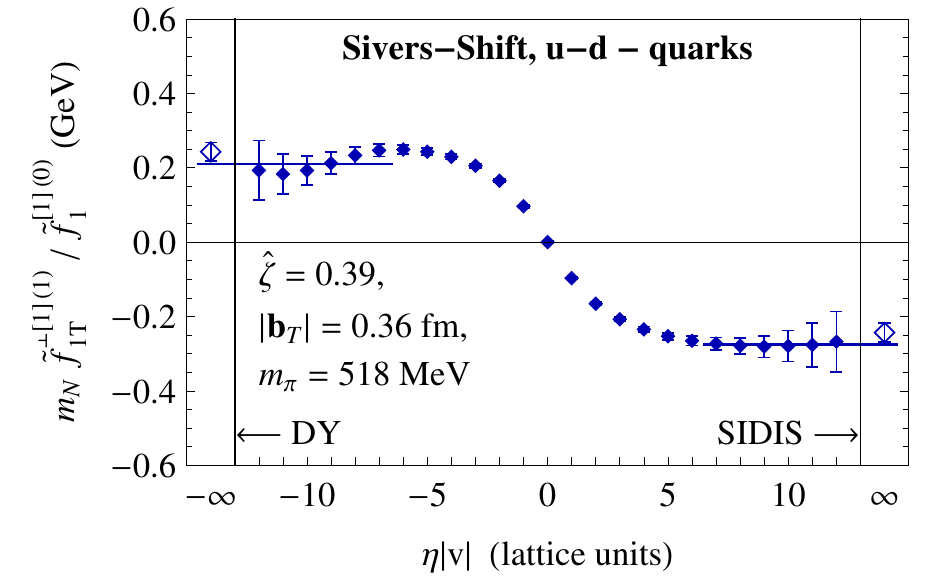}\hspace{1cm}
	\includegraphics[width=0.45\columnwidth]{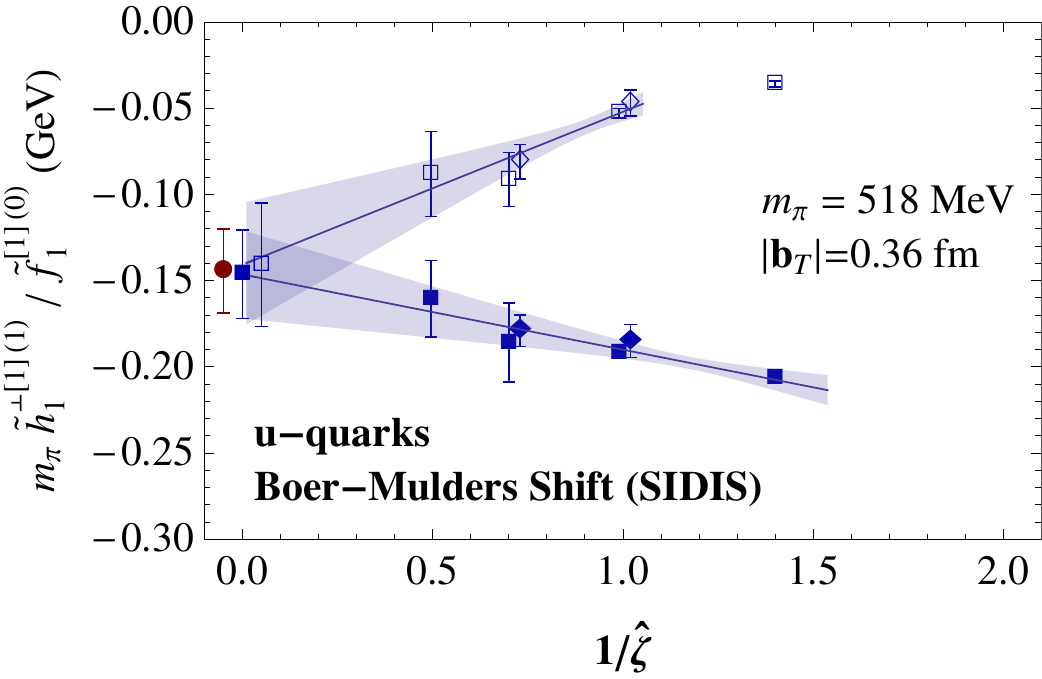}
	\caption{Left: Proton Sivers shift as a function of staple length, $\eta$, for fixed
		staple width, $b_T, $ and rapidity (Collins-Soper) parameter, $\hat{\zeta }$;
		$\eta \rightarrow \infty $ defines the SIDIS limit \cite{Musch:2011er}.
		Right: Extrapolation of the SIDIS limit data for the pion Boer-Mulders
		shift to the physical limit of large $\hat{\zeta }$
		at fixed $b_T $ \cite{Engelhardt:2015xja}. Open symbols represent a partial
		contribution that dominates at large $\hat{\zeta } $, providing further
		insight into the approach to the asymptotic regime.}
	\label{fig_sidis}
\end{figure}

To complement and support these efforts, a sustained
project to calculate TMD observables within LQCD was initiated
and developed by USQCD collab members and their collaborators in Refs.~\cite{Hagler:2009mb,Musch:2010ka,Musch:2011er,Engelhardt:2015xja,Yoon:2017qzo,Engelhardt:2017miy}.
TMDs are formally defined through matrix elements of a bilocal quark
operator in which the quark fields are connected through a gauge link
along a staple-shaped path. Building on the preliminary investigations of Refs.~\cite{Hagler:2009mb,Musch:2010ka}, the first full calculation of TMD
observables using staple-shaped gauge links was performed in
\cite{Musch:2011er}, obtaining results on the Sivers and Boer-Mulders
shifts, a worm-gear shift, and the generalized transversity.
Fig.~\ref{fig_sidis} (left) displays a  result for the Sivers
shift, exhibiting its T-odd character and the SIDIS and DY limits
achieved for asymptotic staple lengths.

Such lattice TMD calculations face several challenges. One such challenge is achieving
the  limit of large rapidity difference between between struck
quark and hadron remnant in a deep inelastic scattering process, which
is encoded in the space-time direction of the staple link. An investigation
of the Boer-Mulders shift in a pion dedicated to elucidating this limit
was reported in Ref.~\cite{Engelhardt:2015xja}. A result from this study is
shown in Fig.~\ref{fig_sidis} (right), demonstrating access to the large
rapidity regime. Another challenge is understanding renormalization,
operator mixing and scaling. Observables such as the Sivers shift are
constructed as ratios in which certain renormalization factors cancel
in continuum QCD; to test whether this pattern persists in the lattice
formulation, a comparison between TMD calculations on clover and domain
wall fermion ensembles at approximately the same pion mass was performed
and reported in Ref.~\cite{Yoon:2017qzo}, corroborating the cancellation of 
renormalization factors expected from continuum QCD.
On the other hand,
in the case of the worm-gear shift, operator mixing is predicted for
clover fermions \cite{Constantinou:2017sej}, which destroys the simple
cancellation in ratios; evidence for this was also seen in the data
collected in \cite{Yoon:2017qzo}.
A preliminary study~\cite{Engelhardt:2015czw} with nearly physical
pions indicates that higher statistical precision is required to impact phenomenology. New efforts
that will be undertaken over the next few years include the use of boosted nucleon sources to access the large
rapidity regime, as well as excited state control through calculation
for a range of source-sink separations.

In addition to the aforementioned calculations, which concentrated on
transverse momentum dependence while integrating over longitudinal
momentum fraction $x$, there are explorations of the
$x$-dependence of the Sivers shift, achieved by adding a
longitudinal separation in the bilocal quark operator defining TMDs.
Furthermore, the generalization
of TMDs to non-zero momentum transfer (GTMDs) was explored in
Ref.~\cite{Engelhardt:2017miy}, with a specific focus on the direct
calculation of quark orbital angular momentum (OAM) in the proton.
Considering non-zero momentum transfer supplements the transverse
momentum information with transverse position information, thus
yielding direct information on OAM (as opposed to indirect access
as $L=J-S$ via Ji's sum rule). Moreover, this approach allows one
to not only determine the Ji OAM, but also the Jaffe-Manohar OAM.

{\it TMDs: Future opportunities:}

The investigations described above provide the necessary foundation for
the controlled, precise prediction of selected TMD observables from
lattice QCD. The chief systematic challenges have been explored,
and a tentative roadmap of incremental refinement of the calculations can
be projected. The use of boosted nucleon sources will allow access to the large
rapidity regime. Discretization effects will need to be quantified
as momenta are increased. This, as well as a quantitative treatment
of the renormalization and QCD evolution of lattice TMD observables,
building on the initial study of Ref.~\cite{Yoon:2017qzo} will necessitate a
sequence of calculations with decreasing lattice spacings.

In assessing the required resources for this program, it should be noted that lattice
TMD calculations are dominated by the cost of the large number of
contractions, as opposed to the cost of the inversions needed to
obtain propagators. The large number of contractions results from the multitude of
staple-shaped gauge link geometries that must be surveyed in order
to perform the necessary extrapolations to long staple length as
well as large rapidity difference between struck quark and hadron
remnant in a deep-inelastic scattering process. 

A further aspect that remains to addressed is the flavor separation
of TMD observables and sea quark effects, as targeted, e.g., by the Fermilab
E-906/SeaQuest experiment. This calls for the evaluation of disconnected
diagram contributions, which hitherto have not been studied in lattice TMD
investigations. Efficient calculation of these diagrams will be possible
with the use of hierarchical probing methods \cite{Stathopoulos:2013aci}.
Future calculations will also need to account for mixing of gluonic
operators with flavor singlet quark operators, which have not been considered.

Besides these improvements of the systematics of lattice TMD
calculations,  the incorporation of further physics objectives is
also planned. To date, calculations have focused on transverse nucleon
polarization, with which one can probe the particularly interesting
Sivers and Boer-Mulders effects. Nonetheless,  the TMDs associated
with longitudinal nucleon polarization are also of interest and include a
second worm-gear function in addition to the one probed with transverse
polarization. TMD calculations with longitudinal polarization are
straightforward to implement in the existing scheme. Furthermore, the
extension of lattice calculations to include the $x$-dependence of TMDs,
already explored for the case of the Sivers shift as noted above, must be
continued to encompass a variety of TMD observables.

In addition, the study of GTMDs, i.e., TMDs in the presence of a
momentum transfer, must be extended beyond the specific case of quark
orbital angular momentum \cite{Engelhardt:2017miy}. For example, spin-orbit correlations of quarks
in the proton can be quantified through the GTMD $G_{11} $ in the
classification scheme of Ref.~\cite{Meissner:2009ww}.
Complementary ways of accessing quark orbital
momentum, e.g., through the twist-3 GTMDs $F_{27} $, $F_{28} $, related
to the GPD $\widetilde{E}_{2T} $, will also be explored  \cite{Meissner:2009ww}.

\subsection{Gluon aspects of hadron structure}

While gluons and the QCD interactions they embody play an essential role in the binding of hadrons, gluon contributions to hadron structure observables are far less well known than their quark analogues. Understanding the role of gluons in hadron structure has become a major goal of experimental facilities, such as COMPASS~\cite{Adare:2014hsq} and STAR~\cite{Djawotho:2013pga}. Furthermore, a primary mission of the proposed Electron-Ion Collider~\cite{Accardi:2012qut,Kalantarians:2014eda}, which is the highest priority for new construction in the NSAC nuclear physics long-range plan~\cite{Geesaman:2015fha}, is to image the gluon structure of hadrons and nuclei. This program will access the three-dimensional gluon structure of the nucleon and allow first measurements of gluon GPDs and TMDs, complimenting significant efforts at RHIC to measure the gluon contribution to the nucleon spin,  potential experiments to study gluon distributions at JLab~\cite{Maxwell:2018gci,Hattawy:2017woc,Dobbs:2017vjw}, and those at the LHC~\cite{Baltz:2007kq}. 
In this light, LQCD calculations of gluon structure quantities have taken on renewed importance. There has been significant progress on this front over the last five years~\cite{Alexandrou:2017oeh,Yang:2016plb,Detmold:2016gpy,Detmold:2017oqb,Winter:2017bfs,Alexandrou:2016ekb}, expanding and building on pioneering studies of the unpolarised gluon structure of the pion and nucleon~\cite{Meyer:2007tm,Horsley:2012pz,Alexandrou:2013tfa,Deka:2013zha} over the last decade. 

\begin{figure}
	\centering
	\includegraphics[width=0.7\linewidth]{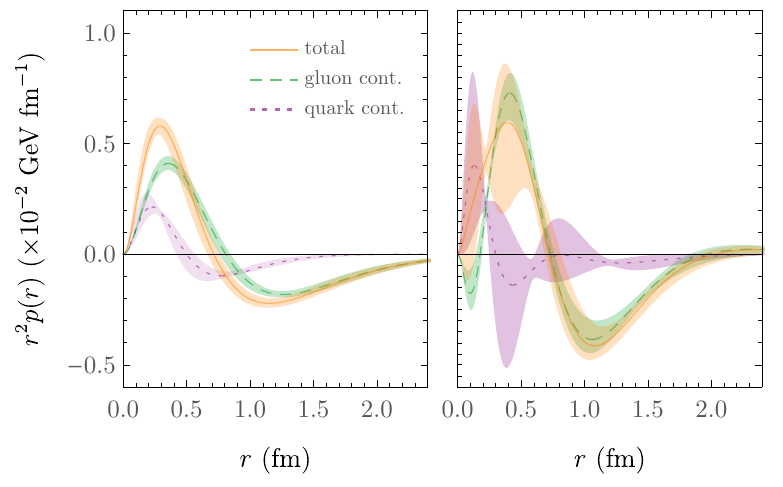}
	\caption{The quark and gluon contributions ot the total pressure distribution in the proton from LQCD. Taken from Ref.~\cite{Shanahan:2018pib}. The left panel corresponds to dipole parameterisations, while the right panel corresponds to the more general $z$-parameterisation form.}
	\label{gluefig}
\end{figure}

In particular, LQCD calculations have provided new insight into the proton spin crisis---the realization that quarks carry only a relatively small fraction of the proton spin---with calculations of the key and poorly-known gluon contributions to the nucleon spin~\cite{Alexandrou:2017oeh,Yang:2016plb}.  As compared to global analyses of polarized parton distributions~\cite{deFlorian:2014yva}, a significantly improved constraint on the total gluon helicity is included in Ref.~\cite{Yang:2016plb}. An important component of these studies is the renormalization of the gluonic operators, which is being achieved using  perturbative~\cite{Glatzmaier:2014sya,Alexandrou:2017oeh} and non-perturbative~\cite{Yang:2018bft} approaches. Complementing this direction, new understanding of the decompositions of the nucleon spin within LQCD has been achieved, giving interpretation to the   orbital angular momentum~\cite{Engelhardt:2017miy}. 
In another impressive success, the gluon contribution to the nucleon's momentum has been resolved at 10\% precision, with the momentum sum rule (including separate determinations of the quark and gluon connected and disconnected contributions) found to be satisfied at quark masses corresponding to the the physical value of the pion mass~\cite{Alexandrou:2017oeh}. Extensions of this work will rival the precision of phenomenological parton distribution fits (e.g., CT14NNLO~\cite{Dulat:2015mca}) in the next few years \cite{Lin:2017snn}.
First calculations of some of the moments of the gluon GPDs~\cite{Diehl:2003ny} that describe the distribution of gluons in hadrons both in the transverse plane and in the longitudinal direction~\cite{Detmold:2016gpy,Detmold:2017oqb} have also been performed, providing  insight into details of the three-dimensional gluon structure of hadrons, albeit without full investigation of systematic uncertainties. 
First calculations of the momentum transfer dependence of the gluon energy-momentum tensor form factors as well as the gluon contributions to the pressure and shear distributions in the proton have recently been performed \cite{Shanahan:2018nnv,Shanahan:2018pib} (see Fig.~\ref{gluefig}). These latter calculations have been combined with recent experimental studies of the quark contributions to the pressure \cite{Burkert:2018bqq}, leading to a first complete determination of this fundamental quantity.
Moreover, aspects of the gluon structure of nuclei have been studied for the first time,  as described in Section~\ref{sec:nuclearstructure}.

{\it Gluon structure: future opportunities}

Exascale computing resources and concurrent algorithm development will facilitate LQCD calculations of static gluon structure quantities with controlled statistical and systematic uncertainties. Calculations on  large lattice volumes that become possible with such resources  will achieve significant precision gains through volume averaging and thereby reduce the gauge noise, which is a statistical challenge for calculations of gluon observables. Nevertheless, these studies face large analysis costs and achieving controlled estimates for non-static quantities requires elimination of excited states, and extrapolation to infinite volume as well as to the continuum limit. 
This becomes especially challenging for large nucleon momenta (as required to extract the $x$-dependence of PDFs, GPDs and TMDs) and in the approach to the continuum limit. 
In the near term, precise calculations of moments of gluon distributions encoding the contribution of gluons to the mass, momentum, and spin of the nucleon and of other hadrons will be refined. In particular, one can expect calculations at quark masses corresponding to the physical pion mass with fully-controlled uncertainties at the level of 2-5\% precision. To achieve this level of systematic control, it is necessary to precisely determine the renormalization factors, including mixing between the gluon observables and the flavor-singlet quark disconnected terms. This carries significant computational cost in its own right.\\

The gluon radius of the nucleon is a quantity as fundamental as the charge radius. Currently, it is not known quantitatively or qualitatively, from experiment or theory, how the charge and gluon radii compare. 
Defined by the slope of the spin-averaged gravitational form factor at zero momentum transfer, the gluon density radius is related via the operator product expansion to matrix elements of the gluon part of the energy-momentum tensor.
The radii and $Q^2$-dependence of the generalized gluon form factors can be calculated using LQCD for both hadrons and light nuclei~\cite{Detmold:2017oqb,Winter:2017bfs}.
On a few-year timescale, fully-controlled calculations of gluon generalized form factors for the nucleon, for low moments and to a scale of several GeV$^2$, can be expected.
From experiment, comparison of nuclear quark and gluon radii will likely be possible through measurements of the parton densities in ${}^4$He at the JLab 12 GeV program~\cite{Hattawy:2017woc}, or from direct measurements of nuclear and nucleon gluon densities using heavy quark production at the planned EIC~\cite{Chudakov:2016otl}. \\

In the longer term, coinciding with the era of sustained exascale computing, one can expect that the $x$-dependence of gluon PDFs and TMDs will be determined from LQCD. Defined on the light-cone, these quantities can not be calculated directly on a Euclidean lattice but can be accessed via rotations to `quasi' or `pseudo' PDFs, matched back to the physical quantities in the large-momentum limit. For the quark PDFs and TMDs these approaches have shown great promise and early success~\cite{Lin:2014zya,Alexandrou:2015rja} (see also Section~\ref{TMDs}). Ultimately, extending these calculations to include gluon distributions will allow a complete decomposition of the three-dimensional quark and gluon structure of the nucleon.

\section{Hadron Spectroscopy}
\label{sec:hadronspectroscopy}
The aim of hadron spectroscopy is to understand the observed experimental spectrum of hadrons in terms of the underlying theory of quarks and gluons, QCD. Traditionally, attempts to decipher the regularities present in the hadron spectrum have focussed on models having only limited connection to QCD; lattice QCD, which offers a first-principles approach to the theory, has now matured to the point where it is vital to efforts to understand excited hadrons.

In broad terms, one aim of the field is to discover how QCD arranges to have such regularity in the excited spectrum of hadrons, where the bulk of observed meson states can be understood as excitations of a $q\bar{q}$ system and baryons as $qqq$, and to understand whether or not there are states dominated by configurations of higher numbers of quarks (tetraquarks, pentaquarks), or configurations featuring only glue (glueballs), or excited glue coupled to quarks (hybrids). These latter possibilities, most of which are not yet unambiguously observed in experiments, come with potential smoking gun signatures of exotic flavor and/or $J^{PC}$ quantum numbers not accessible to a simple $q\bar{q}$ system~\cite{Meyer:2015eta} ($J$, $P$, $C$ refer to the total angular momentum, parity and charge conjugation properties respectively). Within the established hadron spectrum there are states which pose longstanding mysteries such as the light scalar mesons, $a_0(980), f_0(980)$, where diverse model-dependent explanations have been proposed that include tetraquark configurations and meson-meson molecular structure. Ultimately, an understanding of such states must come from QCD.

Our understanding of the excited hadron spectrum continues to be refined through data obtained in contemporary experimental programs (such as COMPASS, GlueX, CLAS12, BES III, LHCb) which are collecting unprecedented statistics with both established and novel production mechanisms. Observations made by these experiments are introducing new mysteries, such as the ``XYZ'' states in the charmonium region which do not fit into the previously successful modelling of charmonium~\cite{Lebed:2016hpi}. Near future experiments like Belle II and PANDA promise continued new information in particular in the bottomonium and charmonium sectors, and LQCD studies of the relevant spectra will continue to play a vital role in the interpretation of the experimental results in the context of QCD.

\subsection{Light hadron spectroscopy}

The lightest hadrons such as the neutron, proton, pion and kaon are stable against decay within QCD, and their mass and other properties can be computed with precision within lattice QCD by controlling the systematic uncertainties introduced through the lattice spacing, lattice volume and choice of quark mass. When the effects of electromagnetism are additionally accounted for, excellent agreement is found between theory and experiment~\cite{Duncan:1996xy, Blum:2007cy, Borsanyi:2014jba, Horsley:2015eaa, Blum:2010ym, Aoki:2012st, deDivitiis:2013xla}.

Unlike these lightest few states, the vast majority of the hadrons which appear in the Particle Data Tables~\cite{Patrignani:2016xqp} are \emph{unstable resonances}, which decay rapidly to lighter hadrons, and whose existence is inferred from enhancements in multi-hadron final states. Decades of accumulated data has led to an experimental spectrum in which each state may be broadly characterized in terms of a mass, a decay width, and branching fractions describing how often the state ends up in each possible decay mode. 

Earlier LQCD calculations considered the excited hadron spectrum in a simplified manner -- in the case of mesons, a large basis of fermion bilinear operators was used to construct a matrix of correlation functions, which was diagonalized to yield a discrete spectrum of excited state energies. The resulting spectra, determined for isospin, $I=1$, $I=0$ and charmonium~\cite{Dudek:2010wm,Dudek:2013yja, Liu:2012ze} show many of the regularities present in the experimental meson spectrum, but in addition show something not yet unambiguously observed, a clear spectrum of \emph{hybrid mesons}, some with exotic values of $J^{PC}$. Corresponding calculations of the baryon spectrum \cite{Dudek:2012ag} show the presence of a spectrum of \emph{hybrid baryons}, and a phenomenology has been built to describe these observations in QCD \cite{Dudek:2011bn}. 

While these calculations, performed at heavier than physical quark masses, give us a tantalizing glimpse of what lattice QCD can tell us about the hadron spectrum, they are not complete in that the hadronic decay physics of the states is not present in any controlled way -- the excited states are appearing as though they were stable states of definite mass rather than as resonances, and the spectrum obtained is at best a guide to the presence of relatively narrow resonances in a given energy region.

\begin{figure}
\includegraphics[width=0.88\textwidth]{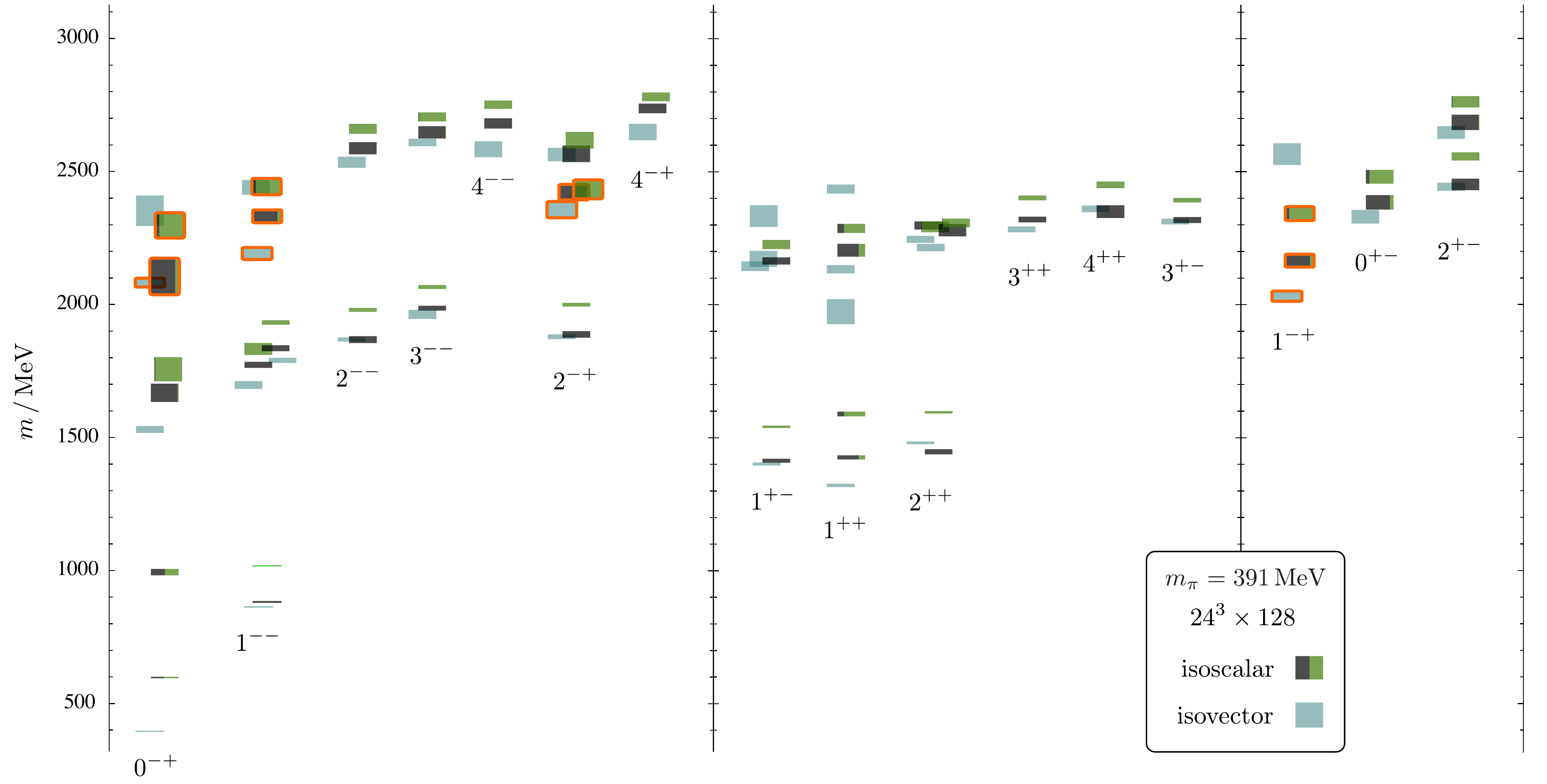}
\caption{The spectrum of excited mesons of various $J^{PC}$ extracted from a lattice QCD calculation with light quark masses such that $m_\pi = 391$ MeV \cite{Dudek:2013yja}. The lightest set of states identified as \emph{hybrid mesons} appear with an orange outline.}
\label{spectrum}
\end{figure}

In order for a QCD calculation to be a faithful reflection the relevant physics, it must be capable of resolving excited hadrons as they truly are, as short-lived resonances, typically decaying to more than one final-state. This necessitates the computation of the energy-dependence of coupled-channel scattering amplitudes, in which the resonances will appear as enhancements. In the past five years, USQCD collaboration members have made significant progress in determining such amplitudes, making use of relations which connect them to the discrete spectrum of eigenstates of quantum field theory in a finite-volume, which can be computed in LQCD
\cite{Luscher:1986pf, Luscher:1990ux, Rummukainen:1995vs, Feng:2004ua, He:2005ey, Bedaque:2004kc, Liu:2005kr, Kim:2005gf, Christ:2005gi, Lage:2009zv, Bernard:2010fp, Fu:2011xz, Leskovec:2012gb,  Briceno:2012yi, Hansen:2012tf, Guo:2012hv, Li:2012bi, Briceno:2013hya, Briceno:2014oea}.

The simplest case is elastic scattering, where in a limited energy region only one hadron-hadron channel is kinematically open. Resonances appearing in elastic scattering include the $\rho$ and the $\sigma$ in $\pi\pi$ scattering, the $K^\star$ in $\pi K$, and the $\Delta$ in $\pi N$, all of which have been considered in LQCD~\cite{
Aoki:2007rd,
Feng:2010es,
Lang:2011mn,
Aoki:2011yj,
Dudek:2012xn,
Pelissier:2012pi,
Wilson:2015dqa,
Bali:2015gji,
Bulava:2016mks,
Guo:2016zos,
Briceno:2016mjc,
Guo:2018zss,
Bali:2015gji, Lang:2012sv, Fu:2012tj, Prelovsek:2013ela,Brett:2018jqw,
Andersen:2017una}.
In the elastic case, the scattering amplitude can be described by a single real energy-dependent parameter, the phase-shift, which has a characteristic rise through $90^\circ$ if a narrow resonance is present. For elastic scattering, there is a simple one-to-one mapping of each discrete energy level in a finite-volume to a value of the elastic scattering phase-shift at that energy (neglecting higher partial waves) \cite{Luscher:1986pf,Luscher:1990ck}. It follows that the lattice calculation is required to have a robust determination of the discrete spectrum of eigenstates, ideally in several lattice volumes. Additionally, the use of moving frames \cite{Rummukainen:1995vs} and/or asymmetrical volumes \cite{Li:2003jn,Detmold:2004qn}, can give access to more energy values which can be used to map out the energy dependence of the phase-shift. To more reliably extract the complete spectrum of eigenstates it proves necessary to go beyond the kind of ``single-hadron-like'' operator basis used in the simplified  spectrum calculations described above, and to also include a set of operators which resemble the relevant hadron-hadron pair undergoing the scattering process.

\begin{figure}
\includegraphics[width=0.44\textwidth]{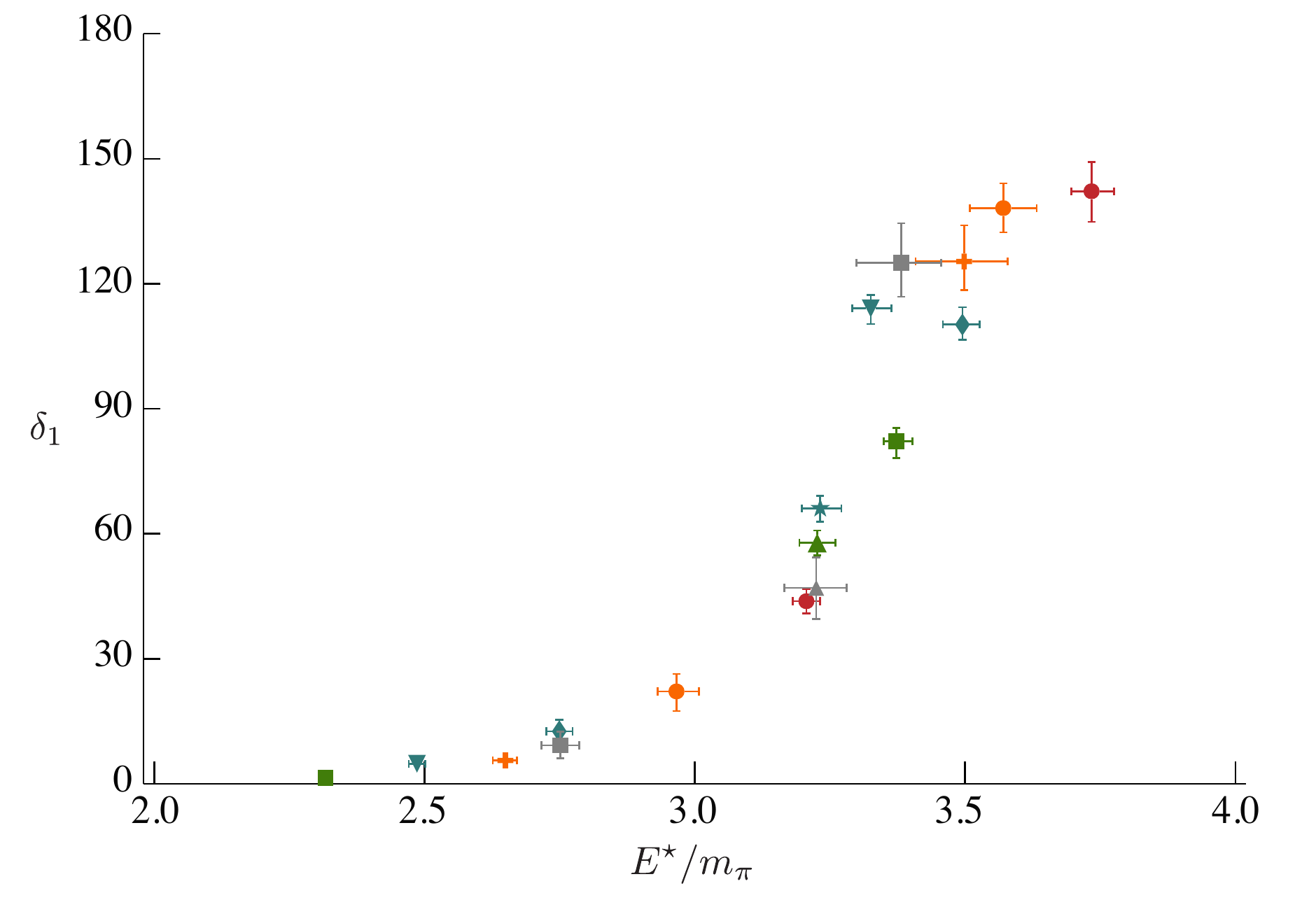}
\includegraphics[width=0.48\textwidth]{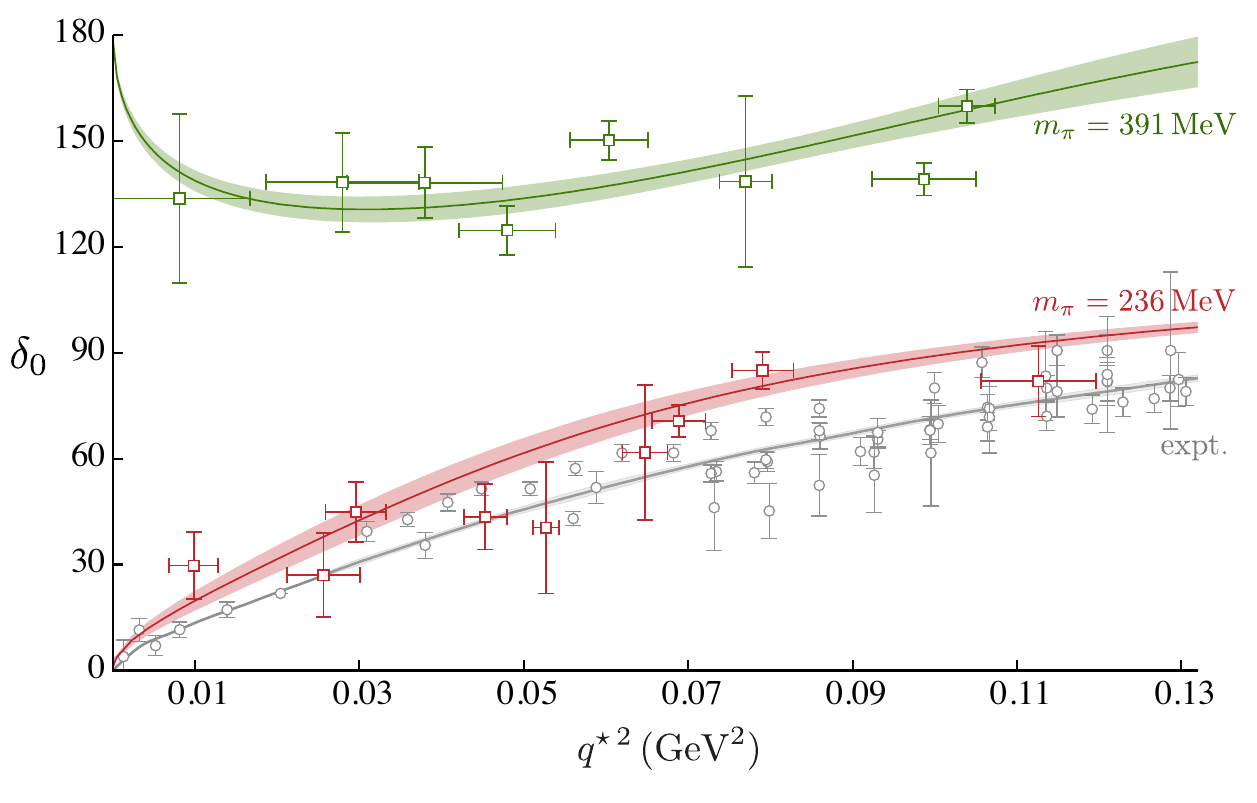}
\caption{$\pi\pi$ elastic scattering phase-shifts from lattice QCD calculations. (a) Isospin=1 $P$-wave computed with $m_\pi \sim 236$ MeV~\cite{Bulava:2016mks} showing the characteristic narrow $\rho$ resonance. (b) Isospin=0 $S$-wave at $m_\pi \sim 391$ MeV and $m_\pi \sim 236$ MeV where the $\sigma$ meson appears as a bound-state, broad resonance respectively~\cite{Briceno:2016mjc}.}
\label{elastic}
\end{figure}

The LQCD technology of operator and correlation function construction has been developed to a state where these elastic scattering calculations are becoming a standard component of the USQCD program and are being pursued by groups around the world
\cite{
Aoki:2007rd,
Feng:2010es,
Lang:2011mn,
Aoki:2011yj,
Dudek:2012xn,
Pelissier:2012pi,
Mohler:2012na, 
Prelovsek:2013cra, 
Lang:2014yfa,
Bali:2015gji,
Lang:2015sba,
Lang:2016jpk,
Bulava:2016mks}. Recent examples are presented in Fig~\ref{elastic} for the case of $\pi\pi$ scattering in $I=1$ and $I=0$, where the very different behavior of the $\rho$ resonance and the $\sigma$ can be observed. LQCD~\cite{Briceno:2016mjc} has shown for the first time in a first-principles approach to QCD, that the $\sigma$ meson evolves from being a broad resonance at light quark masses~\cite{Guo:2018zss}, into a stable bound-state below the $\pi\pi$ threshold. 

Going beyond the simplest case of elastic scattering, resonances will appear in the \emph{coupled-channel $S$-matrix}. When more than one hadron-hadron channel is open, the lattice spectrum calculations require extension of the operator basis to include hadron-hadron operators of the relevant species, but the analysis to turn this spectrum into information about scattering is less straightforward, since there can no longer be a one-to-one mapping of any given energy level into the multiple unknowns of a coupled-channel scattering matrix at that energy \cite{Luscher:1986pf}. A successful approach \cite{Dudek:2014qha, Wilson:2014cna, Wilson:2015dqa, Moir:2016srx, Dudek:2016cru, Briceno:2017qmb} has been to parameterize the energy-dependence of coupled-channel amplitudes, and to use very many discrete energy levels in multiple volumes and/or moving frames to constrain the free parameters. Potential bias introduced by explicit choice of parameterization can be reduced by considering a range of forms, and exploring to what extent the best-fit amplitudes vary with parameterization choice~\cite{Dudek:2014qha, Wilson:2014cna, Wilson:2015dqa, Moir:2016srx, Dudek:2016cru, Briceno:2017qmb}. 

\begin{figure}
\includegraphics[width=0.8\textwidth]{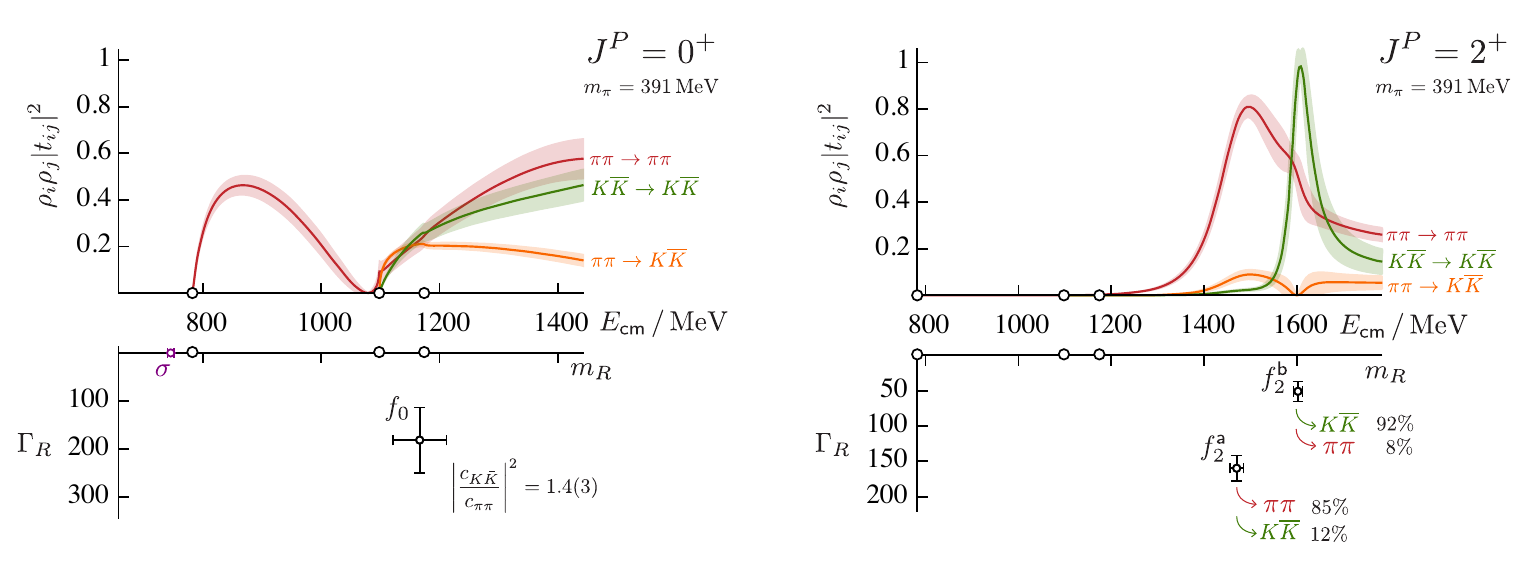}
\caption{$\pi\pi, K\overline{K}$ coupled-channel scattering amplitudes in two partial-waves determined in a lattice QCD calculation with $m_\pi \sim 400$ MeV~\cite{Briceno:2017qmb}. (a) $J^P=0^+$ sector found to contain a bound-state $\sigma$, but also a resonance pole near the $K\overline{K}$ threshold, having strong coupling to both $K\overline{K}$ and $\pi\pi$ that may be associated with the experimental $f_0(980)$ meson. (b) $J^P=2^+$ sector found to contain two narrow tensor resonances.}
\label{f0f2}
\end{figure}

Having explicit analytic forms for the amplitudes has the advantage that it becomes possible to determine resonance properties  by analytically continuing the parametrized amplitudes into the complex energy plane, with resonances appearing as pole singularities, and where the couplings of resonance states to their decay channels can be obtained from the pole residues. The real and imaginary parts of the pole position can be identified with the mass and total width of the resonance, and the couplings are related to the decay branching fractions.
This methodology was recently used to find low-lying scalar and tensor resonances in the coupled $\pi\pi, K\overline{K}, \eta\eta$ system with $I=0$. In Ref.~\cite{Briceno:2017qmb}, a calculation with quark masses corresponding to a pion mass $\sim$ 400 MeV was presented where excited state spectra were extracted from variational analysis of correlation matrices computed in three lattice volumes, in a range of moving frames. The resulting energies were used to constrain the various amplitudes shown in Fig.~\ref{f0f2}. The scalar amplitude has a highly non-trivial behavior in which a bound-state lying below $\pi\pi$ threshold interferes with an $f_0(980)$-like resonance singularity lying close to the $K\overline{K}$ threshold, leading to a {dip} in the $\pi\pi \to \pi \pi$ amplitude that is analogous to a feature seen in the experimental amplitude. The resonance is found to have roughly equal coupling strength to $\pi\pi$ and $K\overline{K}$. The tensor amplitude is quite different, being much closer to our expectations for straightforward resonant enhancements, with two clear peaks corresponding to two pole singularities, one coupled dominantly to $\pi\pi$ and the other to $K\overline{K}$; numerical estimates are determined for the branching fractions from the pole residues. These two resonances closely resemble the experimentally well established $f_2(1270), f_2'(1525)$ states.
This example illustrates the highly non-trivial dynamics that can arise in hadron-hadron scattering from the non-perturbative dynamics of QCD, and LQCD is for the first time providing us a methodology to explore this dynamics without recourse to approximations or assumptions whose justification may not be clear. 

\begin{SCfigure}
\includegraphics[width=0.45\textwidth]{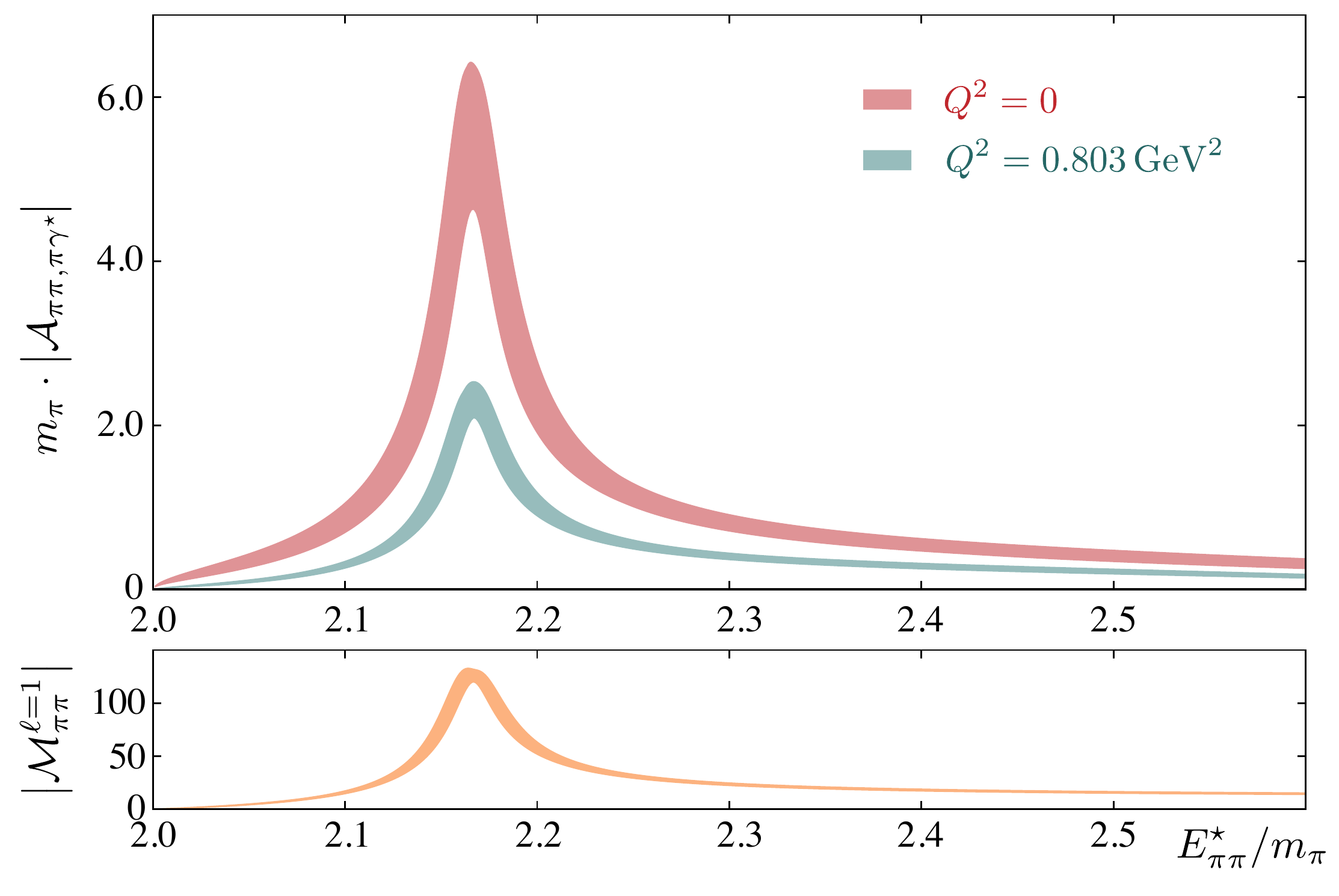}
\caption{Amplitude for the process ${\pi \gamma^\star \to \pi\pi}$ in $P$-wave computed in lattice QCD with $m_\pi \sim 400$ MeV~\cite{Briceno:2016kkp,Briceno:2015dca}. Shown for two values of the photon virtuality, $Q^2$, and also shown the corresponding amplitude for $\pi\pi \to \pi \pi$ indicating that the $\rho$ resonance is contributing to both processes.  }
\label{rhopigamma}
\end{SCfigure}

A new generation of experiments are studying hadron spectroscopy using novel production mechanisms -- an example being the GlueX experiment at Jefferson Lab, which is producing meson resonances using a photon beam, with a particular focus being the search for exotic $J^{PC}$  mesons which may have an explanation as \emph{hybrids} featuring an excitation of the gluonic field. The anticipated huge data set from this experiment motivates study of the coupling of excited hadrons to photons, and in recent years we have seen the development of a formalism to extract the relevant amplitudes from finite-volume LQCD calculations \cite{Briceno:2014uqa,Briceno:2015csa,Briceno:2015tza}. Indeed, the first explicit calculation \cite{Briceno:2015dca,Briceno:2016kkp} computed three-point vector current correlation functions corresponding to the quantum numbers of the process $\gamma^\star \pi \to \pi \pi$ with $J^P=1^-$, in which the $\rho$ resonance is expected to appear. The results of this first calculation at $m_\pi \sim 400$ MeV are presented in Fig.~\ref{rhopigamma} where the effect of the $\rho$ resonance in the electromagnetic transition amplitude can be clearly observed, and where the dependence on the virtuality of current can be used to determine the \emph{transition form-factor} of the unstable $\rho$ resonance (see also \cite{Alexandrou:2018jbt}). The formalism for the analysis of  $e^+ e^-$ annihilation to meson-meson final-states through a photon has also been applied to $\rho \to \pi\pi$ decays~\cite{Feng:2014gba}.

\vspace{5mm}
Over the past ten years LQCD has transformed theoretical hadron spectroscopy, moving it from being dominated by model calculations, which while useful, had only a limited connection to QCD, to being directly connected with QCD with only controlled approximations made. Initial successes mapping out the highly excited spectrum of mesons and baryons, while excluding their decay properties, led to answers to longstanding questions about the role of excited gluonic fields in the hadron spectrum. More recently the field has begun computing excited hadrons as they truly are, as unstable resonances in hadron scattering amplitudes, firstly for the simple case of elastic resonances, and then for resonances which can decay into two or three different final states. The coupling of unstable resonances to external currents is now accessible in lattice QCD calculations, in some cases this provides an observable which can be compared to experiment, and in others a set of form-factors which can be used to build a space-time picture of the distribution of constituents within the unstable hadrons, potentially allowing for a validation of model-dependent claims that some states are e.g. diffuse meson-meson molecules. We are only beginning to see the possibilities of using lattice QCD to understand hadron spectroscopy, as has been highlighted in the 2015 NSAC Long Range Plan for Nuclear Science \cite{Geesaman:2015fha}.

This physics progress comes about because of significant technical advances that have been made by USQCD in all aspects of LQCD calculations (see the  companion whitepaper of computational LQCD \cite{Joo:2018qcd}). Operators have been developed which overlap efficiently with the eigenstates of QCD in a finite-volume, either as ``single-hadron'' operators or as constructions which resemble a pair of hadrons and which respect the cubic symmetry of the lattice. To compute the required correlation functions, diagrams featuring quark-antiquark ``annihilation'' lines are required, and techniques such as \emph{distillation}~\cite{Peardon:2009gh} and stochastic variants~\cite{Morningstar:2011ka} have rendered these, which were traditionally considered extremely challenging, now quite run-of-the-mill. The use of anisotropic lattices, in which the lattice spacing in the time direction is somewhat finer than in the spatial directions, has allowed for rather precise determinations of discrete energy spectra, through fine resolution of the time dependence of correlation functions. Anisotropic lattices have reduced the computational cost of such studies relative to working on fine isotropic lattices, particularly given the need to use relatively large volumes. The need to compute large matrices of correlation functions in order to accurately determine the discrete excited spectrum of QCD eigenstates, has had the consequence that the computation cost of the final stage of the lattice calculation, in which the correlation functions are constructed, has become a significant portion of the total computational budget. Much ongoing effort is devoted to identifying methods to reduce this cost.

\paragraph{Future opportunities}


Looking forward, we expect to see much more progress in understanding the light hadron spectrum using lattice QCD methods. Early targets will see the application of established two-body techniques to channels that have not previously been explored. Proposed calculations include some to study the experimentally established axial meson resonances like $a_1$, $b_1$, which are known to have dominant decays to $\pi \rho$, $\pi \omega$. At heavier than physical light quark masses, the vector mesons in the decay are stable, and the decay is two-body. There are still novel challenges here associated with the additional spin degree-of-freedom provided by the vector: for example, the $J^{PC}=1^{+\pm}$ quantum numbers of the axial mesons can be constructed with either an $S$-wave or a $D$-wave between the pseudoscalar and the vector mesons. In a non-resonant case of $\pi \rho$ scattering with $I=2$ it has been shown that the relative strength of these two channels, and the mixing between them, can be determined in lattice QCD calculations~\cite{Woss:2018irj}. These techniques, once established for the axial meson resonances can be extended to other $J^{PC}$; an important case is the exotic $J^{PC}$ sector in which \emph{hybrid mesons} are predicted to appear. The larger mass of these resonances is such that several decay channels are kinematically accessible. The aim of the first calculations will be to predict some properties of these states in advance of the search within the GlueX data set, and in particular to have first estimates for the mass, total decay width, and the branching fractions to the various final states. This can be used to offer guidance to experiments like GlueX which have to select a particular set of final state particles for analysis when searching for resonances.

While the expectation is for calculations to progressively be done at lighter and eventually physical light quark masses, in the short-term, some calculations at heavier than physical quark masses will continue to be warranted. By increasing the light quark mass, pions become heavier, and three-meson thresholds correspondingly lie higher, providing a larger energy region over which the unique and well-studied two-body finite-volume formalism can be applied rigorously. 
At present, the absence of a complete formalism to describe three-hadron scattering in a finite-volume is an important  restriction. It is clear that this must be remedied if calculations are to proceed at lighter quark masses, where the bulk of resonances lie above at least one three-hadron threshold. On this front, a significant formal effort is underway \cite{Polejaeva:2012ut,Hansen:2015zga,Hansen:2014eka,Briceno:2018aml,Doring:2018xxx} using a number of different approaches, and there is work ongoing to understand the commonalities in the results. On the practical lattice computational side, the extension of previous calculations is relatively straightforward -- three-hadron-like operators can be constructed using the same techniques used to combine single hadrons into two-hadron operators, and approaches like distillation \cite{Peardon:2009gh} allow for the relevant correlation functions to be computed without any additional computation of propagator objects. The increased number of quark fields involved will naturally lead to a combinatoric increase in contraction costs, and algorithmic improvements under the LQCD Exascale Computing Project and the LQCD SciDAC-4 project are being explored to reduce these costs.

While the development of a rigorous three-body (and higher) formalism is vital to have confidence in the calculations of high-lying resonances, it is likely that explicit calculations will show simpler behavior corresponding to quasi-two-body decays in many cases. Experimentally resonances appearing in three-body and higher multiplicity final state are observed to dominantly proceed through intermediate two-body states featuring isobar resonances which subsequently decay, e.g. $a_1 \to \rho \pi \to (\pi\pi) \pi$. It may eventually prove possible to make use of this isobar dominance to simplify somewhat the analysis of finite-volume spectra in energy regions in which three hadron and even higher multiplicity final states are kinematically accessible.

Building on the first successful calculations involving currents coupled to resonances, we will see extensions to other resonant states. Transition form-factors evaluated for photons with zero momentum transfer control the rate of photoproduction at GlueX -- first calculations (even for unphysically heavy quark masses) of established mesons can be compared to the first round of analysis of the GlueX data set, and prediction estimates made for the exotic $J^{PC}$ state production rates. Beyond electromagnetism, we will see calculations of light quark resonances appearing in weak heavy-flavor decays. This includes the flavor-changing neutral-current process $B \to \ell^+\ell^- \!\!\!\! \underset{\;\;\;\;\;\hookrightarrow K\pi}{K^*}$, in which there are currently tensions between theory and experiment that hint at physics beyond the Standard Model, and the charged-current decay $B \to \ell^-\bar{\nu}  \!\!\!\!\!\!\!\!\;\;\underset{\;\;\;\;\;\;\hookrightarrow \pi\pi}{\rho}$, which can provide new information on the $|V_{ub}|$ puzzle. More detailed discussions of these weak decays can be found in
the accompanying whitepaper on quark and lepton flavor physics \cite{Lehner:2018qcd}.

As described in Section~\ref{sec:hadronstructure}, there are opportunities to use the techniques developed for spectroscopic studies of resonances to investigate their three-dimensional gluon structure described by gluon GPDs and TMD. These quantities may provide insight from QCD into details of the nature of exotic states. These calculations are extremely demanding computationally and will also require continued theoretical development.

\subsection{Heavy quarks and the XYZ states}

Since the  observation of the $X(3872)$ in 2003 \cite{Choi:2003ue}, an ever growing family of unexpected enhancements in the experimental studies of charmonium region have been seen, known colloquially as the ``XYZ'' states. These enhancements, if interpreted as resonances, typically lie outside the previously successful picture of charmonium in terms of $c\bar{c}$ bound-states, sometimes in extreme ways. For example the $Z_c$ enhancements observed in final states like $J/\psi \, \pi^+$ are \emph{charged}, and it is argued must have minimal quark content $c\bar{c} u \bar{d}$. Further discoveries and refined measurements of the properties of observed states continue in earnest at facilities like LHCb and BES III, with further extension into the bottomonium sector expected at Belle II.

Within the charm sector, the LQCD methods described above can be brought to bear on the question of flavor exotics and the other excess XYZ states. There have been suggestions that at least some of the observed experimental enhancements arise due to the kinematics of the three-body production process (e.g. $e^+e^- \to \pi \; (\pi J/\psi) $ or $B \to K \, (\psi' \pi)$), rather than being due to a true two-body resonance \cite{Szczepaniak:2015eza}. Lattice calculations have the advantage here that in order to determine the resonant content, they are not restricted to studying particular higher-multiplicity production processes, but rather they can compute the two-body scattering amplitude directly, removing the effect of any kinematic singularities particular to the production mechanism.

The techniques for determination of coupled-channel scattering matrices pioneered in the light-quark sector and described in the previous section can be applied for heavy quarks. The calculations are somewhat more technically challenging as the small spin-splitting between $D$ and $D^*$ mesons, and the lightness of the $\pi$ compared to the energy gap between the $J/\psi$ and the relevant excited states, means that there are typically several kinematically accessible channels which must be considered.

The calculation of the radiative decay of the XYZ states can address the speculations of the `XYZ'-s are `molecular' in origin. First calculations could target the open-charm systems as well as the more challenging $I=0$ $D\bar{D}$ decays.
There are hints already that some of the experimentally observed enhancements may not have a resonant origin. In Ref.~\cite{Cheung:2017tnt} (see also \cite{Prelovsek:2014swa}), a lattice calculation of the spectrum of states with the quantum numbers of the $Z_c(3900)$, using a large basis of operators containing many resembling the expected finite-volume meson-meson states as well as several having tetraquark-like structure, showed no significant deviations from the spectrum expected if interactions are only weak, and no resonance is present. 

While the first LQCD studies suggest that double charm and hidden charm tetraquarks do not appear as entities in the spectrum, there is significant evidence from lattice calculations that double beauty tetraquarks are actually {bound}~\cite{Francis:2016hui, Junnarkar:2018twb, Leskovec:2019ioa} (see also Refs.~\cite{Hughes:2017xie, Francis:2018jyb}). Such states, if they can be produced experimentally, would be observed through their weak decay. Further LQCD calculations, utilizing the diverse operator bases already shown to be capable of  reliably extracting the complete low-energy spectrum, are warranted to investigate systematics and determine the properties of these states with higher precision.

The spectrum and dynamics of hadrons containing heavy quarks are constrained by approximate heavy-quark flavor and spin symmetries
\cite{Korner:1994nh, Manohar:2000dt}. A particularly interesting symmetry emerges for doubly heavy baryons and doubly heavy tetraquarks: in the large-mass limit, the two heavy quarks
are expected to form a point-like diquark that acts like a single heavy antiquark, and the light degrees of freedom behave as in a
singly-heavy hadron \cite{Carlson:1987hh, Savage:1990di, Manohar:1992nd, Brambilla:2005yk, Eichten:2017ffp}. With the current operation of the LHC, charm and bottom baryons are being produced in unprecedented quantities.
This has led to several discoveries in the last few years \cite{Chatrchyan:2012ni, Aaij:2012da, Aaij:2014yka, Aaij:2016jnn, Aaij:2017ueg,  Aaij:2017vbw, Aaij:2017nav, Aaij:2018yqz, Aaij:2018tnn}, with many more expected in the future. LQCD
can predict the masses, can help assign $J^P$ quantum numbers, and can also provide information on the structure and decay rates~\cite{Brown:2014ena, Padmanath:2015jea, Bali:2015lka, Can:2015exa, Alexandrou:2016xok, Bahtiyar:2018vub, Woloshyn:2016pid, Alexandrou:2017xwd, Mathur:2018epb, Mathur:2018rwu}. Including the effects of electromagnetism and isospin breaking even allows estimates of charge splittings for stable states~\cite{Borsanyi:2014jba}.

The LHCb collaboration has reported the observation of three narrow $J/\psi\: p$ pentaquark resonances, $P_c(4312)$, $P_c(4440)$, and $P_c(4457)$, in $\Lambda_b \to J/\psi\,p\,K$ decays \cite{Aaij:2015tga, Aaij:2019vzc}.
Studying these resonances on the lattice is challenging due to the many open channels, including channels with more than two hadrons. Charmonium-nucleon
interactions have been investigated in lattice QCD at low energies \cite{Yokokawa:2006td, Liu:2008rza, Kawanai:2010ev, Beane:2014sda}, and recently also in the $P_c$ energy region \cite{Skerbis:2018lew}. The interactions near threshold were found to be slightly attractive, with an increasing attraction at unphysically heavy up and down quark quark masses, where bound states were seen \cite{Beane:2014sda}. The recent study of charmonium-nucleon interactions at higher energies \cite{Skerbis:2018lew} did not find any $P_c$ resonance, but the inclusion of additional channels (such as $\Sigma_c^+ \bar{D}^{0(*)}$) is expected to be important.

\section{Nuclear Spectroscopy, Interactions and Structure}
\label{sec:nuclear}

 Beyond the physics of single hadrons described in the previous sections, the complexity of the nuclear landscape emerges from QCD and the other forces of the SM.  From the point of view of QCD, this emergence is an interesting phenomena, with the various effective degrees of freedom in nuclei (nucleons, $\alpha$ clustering, shells, resonances) all being extremely non-trivial consequences of QCD dynamics that beg for explanation. The first steps in addressing this complexity from LQCD have been made over the last decade, and we anticipate that LQCD will become an increasingly important part of nuclear theory in the coming years. Since the SM forms the foundation of nuclear physics, LQCD can be used to study the forces that bind nucleons into nuclei and govern their interaction, as well as to investigate how nuclear systems interact with external electroweak probes and possible physics beyond the Standard Model.
While there has been remarkable progress in this area over the last decade, it is clearly an area where major opportunities for new developments exist and major challenges await.

\subsection{Nuclear spectroscopy}

Determining the ground state energies and bindings of light nuclei is a central challenge for LQCD in nuclear physics. The very first LQCD calculations of nuclei (objects with baryon number greater than one) are less than a decade old and significant advances in the study of nuclear systems have occurred over the last five years. Although no bound states were determined in earlier studies of baryon-baryon systems \cite{Yamazaki:2009ua,Beane:2006gf,Beane:2006mx,Beane:2009gs,Beane:2009kya}, these works  developed the necessary contraction and analysis techniques for efficient study of two-body systems. 
The first calculations of bound systems with baryon number $A\ge2$ were of the doubly-strange $H$-dibaryon system at unphysical quark masses \cite{Beane:2010hg,Inoue:2010es,Beane:2011xf}. 
Studies of the two-nucleon channels \cite{Beane:2011iw} and other exotic channels were performed subsequently \cite{Berkowitz:2015eaa,Francis:2018qch,Wagman:2017tmp}. These studies used the same L\"uscher method discussed above in Section \ref{sec:hadronspectroscopy}, converting finite volume energy eigenvalues into determinations of infinite volume bound state pole positions. Bound states have also been found using the HAL potential method \cite{Ishii:2006ec} based on Refs.~\cite{Luscher:1986pf,Lin:2001ek}, although issues with the validity of current applications of the method have been raised  \cite{Detmold:2007wk,Birse:2012ph,Yamazaki:2018qut,Yamazaki:2017gjl,Namekawa:2017sxs} and it is only recently that systematics have begun to be addressed in this method \cite{Kawai:2017goq}.
A series of studies of systems of many mesons \cite{Beane:2007es,Detmold:2008yn,Detmold:2011kw} allowed the extraction of a three-particle interaction from LQCD for the first time.
Calculations of systems up to atomic number $A=4$  have followed \cite{Beane:2012vq,Yamazaki:2012hi,Yamazaki:2015asa}, with almost physical quark mass calculations being currently performed by the PACS-CS collaboration \cite{Yamazaki:2015asa}. The authors of Refs.~\cite{Iritani:2018zbt} have suggested that systematic issues exist in the extractions of energies in many of these studies. While such issues can potentially exist they require careful investigation on a case-by-case basis and many aspects of the criticism are refuted for particular calculations in Refs.~\cite{Beane:2017edf,Namekawa:2017sxs,Yamazaki:2017jfh}.

Spectroscopy of nuclear systems is particularly challenging for LQCD for multiple reasons, some physical and others technical. The first challenge stems from the fact that the physics of nuclei is complicated, with low-energy excitations possible through many different mechanisms; understanding even the simplest aspects completely requires precise control of spectroscopy. Existing studies are at some level saved by the finite lattice volumes and the heavier than physical quark masses that were used which lead to a simpler spectrum. However future calculations in large volumes and with physical quark masses must confront these issues (Ref.~\cite{Beane:2012vq} highlights just how formidable this challenge is).
A second challenge arises from the Monte-Carlo techniques used for LQCD calculations. As emphasised by Parisi and Lepage \cite{Lepage:1989hd,Parisi:1983ae,Hamber:1983vu}, single baryon correlation functions exhibit a signal-to-noise ratio that degrades exponentially with the temporal separation. For nuclear systems, the problem only becomes more challenging \cite{Beane:2009kya,Beane:2009gs} and extraction of the eigenenergies is consequently difficult. A number of methods have been developed that aim to ameliorate this issue, either defining better-behaved estimators \cite{Beane:2014oea,Wagman:2017gqi,Wagman:2017xfh,Wagman:2016bam,Detmold:2018eqd} or new analysis strategies that optimize the ratio of signal to noise \cite{Detmold:2014hla}. None of these methods has completely solved the signal-to-noise problem, but they have proved sufficient for studies of the lightest few nuclei. 
Finally, at least naively, the complexity of contractions grows 
factorially with the system size; calculations for $^4$He are at first sight $6!6!/2\sim 260,000$ times  more difficult than for a proton. Efforts to reduce these costs via construction of enumerative \cite{Doi:2012xd,Gunther:2013xj} and recursive \cite{Detmold:2010au,Detmold:2012eu} algorithms have enabled the progress described above. 
Given the exponentially hard nature of these challenges, new techniques making use of machine learning and also quantum information science will potentially have transformative  impact 
in  many body lattice QCD in particular. Some new directions are discussed in the companion White Paper on Computational Lattice Field Theory \cite{Joo:2018qcd}.

There are many opportunities for increased effort in this area as well as many technical challenges that exist in studying larger nuclei and performing calculations at the physical quark masses. 
Extending existing calculations to even moderately larger $A$ will have significant impact as nuclei that require $p$-shell configurations become accessible. These systems depend on more complicated aspects of the nuclear forces than the $A\le4$ nuclei that have been studied and new lattice calculations will constrain different spin and isospin components of these forces. Larger nuclei also exhibit interesting collective effects such as halo structures (eg,  ${}^{6,8}$He), cluster structures (eg ${}^{8}$Be, ${}^{12}$C) and deformations that would be very instructive to see emerge from LQCD calculations.  Extension of the current calculations to excited states of nuclear systems will also provide renewed insight into the nature of nuclei, but will require exascale computing and advanced variational techniques such as those discussed in Section \ref{sec:hadronspectroscopy}.
By using a large basis of operators of different symmetries and structures, this would allow a detailed understanding the nature of these excitations and the origin of collectivity in nuclei. 

As well as efforts towards controlled calculations of nuclear physics to understand and interpret experiment, LQCD offers the possibility of investigating nuclei away from the physical quark masses, or for different gauge and fermion content of the theory, as an intellectual pursuit of its own. These systems cannot be studied in experiment, but can provide a broader perspective on the nature of gauge filed theories and promise concrete answers to questions related to the naturalness of nuclear physics \cite{Orginos:2015aya,Epelbaum:2012iu,Berengut:2013nh,CarrilloSerrano:2012ja}. With the  possibility of strongly interacting gauge theories other than QCD occurring in the context of dark matter and hidden valley models, it is also interesting to understand how ubiquitous nuclear physics is and whether there are QCD-like theories that do not exhibit nuclear physics.  In a first step in this direction,  Refs.~\cite{Detmold:2014qqa,Detmold:2014kba}  consider $N_f=N_c=2$ QCD and examine the nuclear physics and phenomenological consequences of  a putative dark matter sector based on this theory. Surprisingly these studies, and studies of $N_c=3$ QCD at unphysical quark masses suggest that the shallow binding of nuclei is a fairly generic phenomenon in theory space. Further investigations of theories such as those with multiple matter representations \cite{Ayyar:2017qdf} may display fundamentally different nuclear phenomena.

\subsection{Nuclear Structure}
\label{sec:nuclearstructure}

Exploration of the structure of the nuclei found in LQCD calculations from the underlying quark and gluon degrees of freedom offers  opportunities for new insight into the complexities of nuclear physics, as well as additional challenges.  Phenomenologically, the spectroscopy and decay patterns of excitations of nuclei have been an important source of structure information. However, interactions of nuclei with electroweak probes have provided the most precise information we have about the internal dynamics of nuclei. 
In particular, the magnetic moments,  higher multipole moments, and polarizabilities  have enabled a static picture of nuclei to be determined. The electromagnetic form factors of nuclei have revealed their charge and current distributions and  have further
developed our understanding of nuclear structure. 
Nuclear parton distributions extracted from deep-inelastic scattering on nuclear targets have provided a different perspective on the partonic substructure of nuclei. Based on this phenomenology, nucleons are seen to be effective degrees of freedom inside nuclei, leading to the success of the phenomenological shell model and many-body approaches based on nucleon degrees of freedom in describing many aspects of nuclear structure. However, there are many ways in which nuclei reveal that non-nucleonic degrees of freedom are important inside the nucleus. The EMC effect \cite{Aubert:1983xm}, discovered in 1983, is perhaps the most striking; it shows that the distribution of quarks and gluons in a nucleus differs at the ${\cal O}(20\%)$ level from the incoherent sum of the distributions in the nucleon. Understanding these and other aspects of nuclear structure from QCD  is an important challenge.

\begin{figure}[!t]
	\centering
	\raisebox{0.1\height}{\includegraphics[width=0.48\columnwidth]{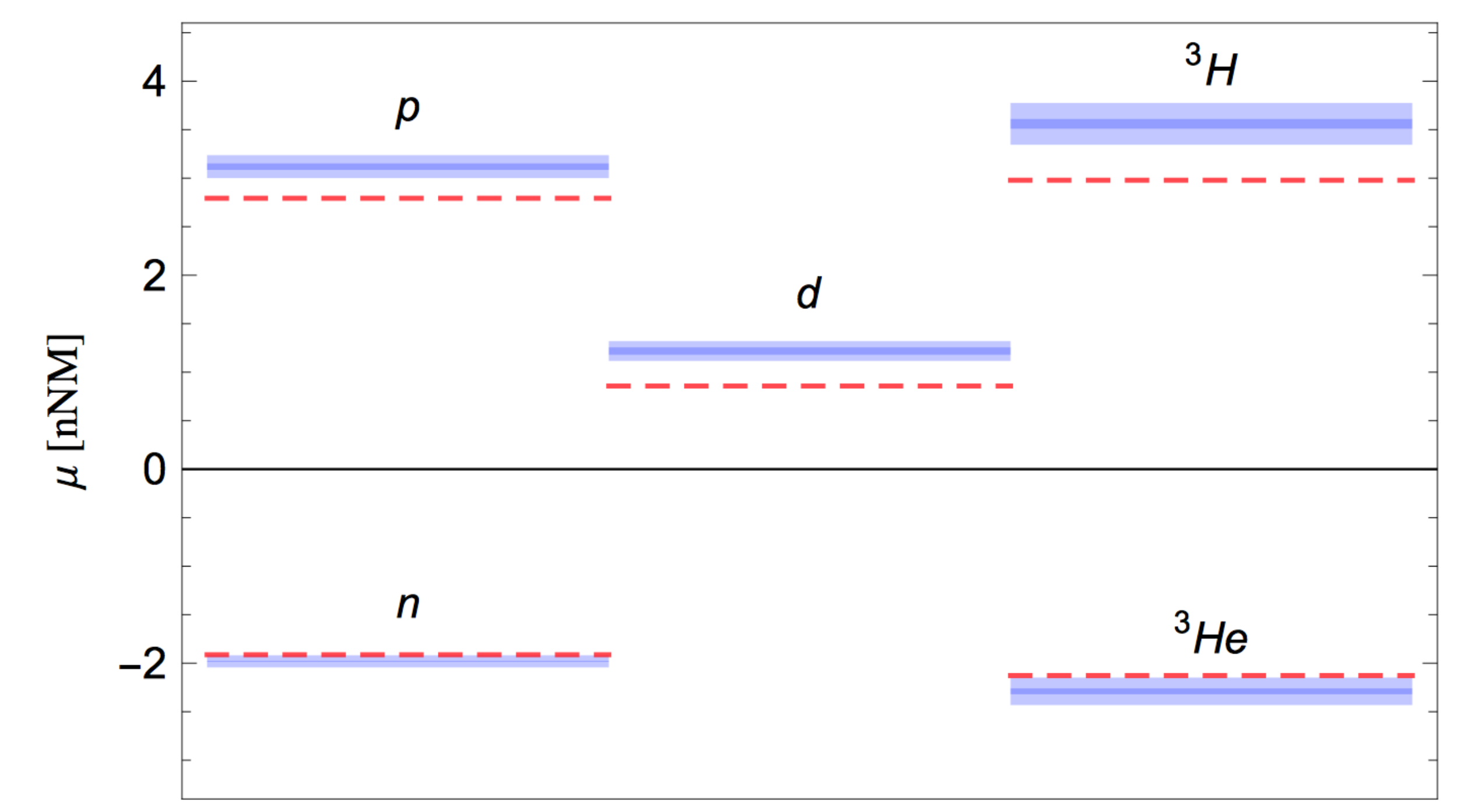}  }
	\includegraphics[width=0.49\columnwidth]{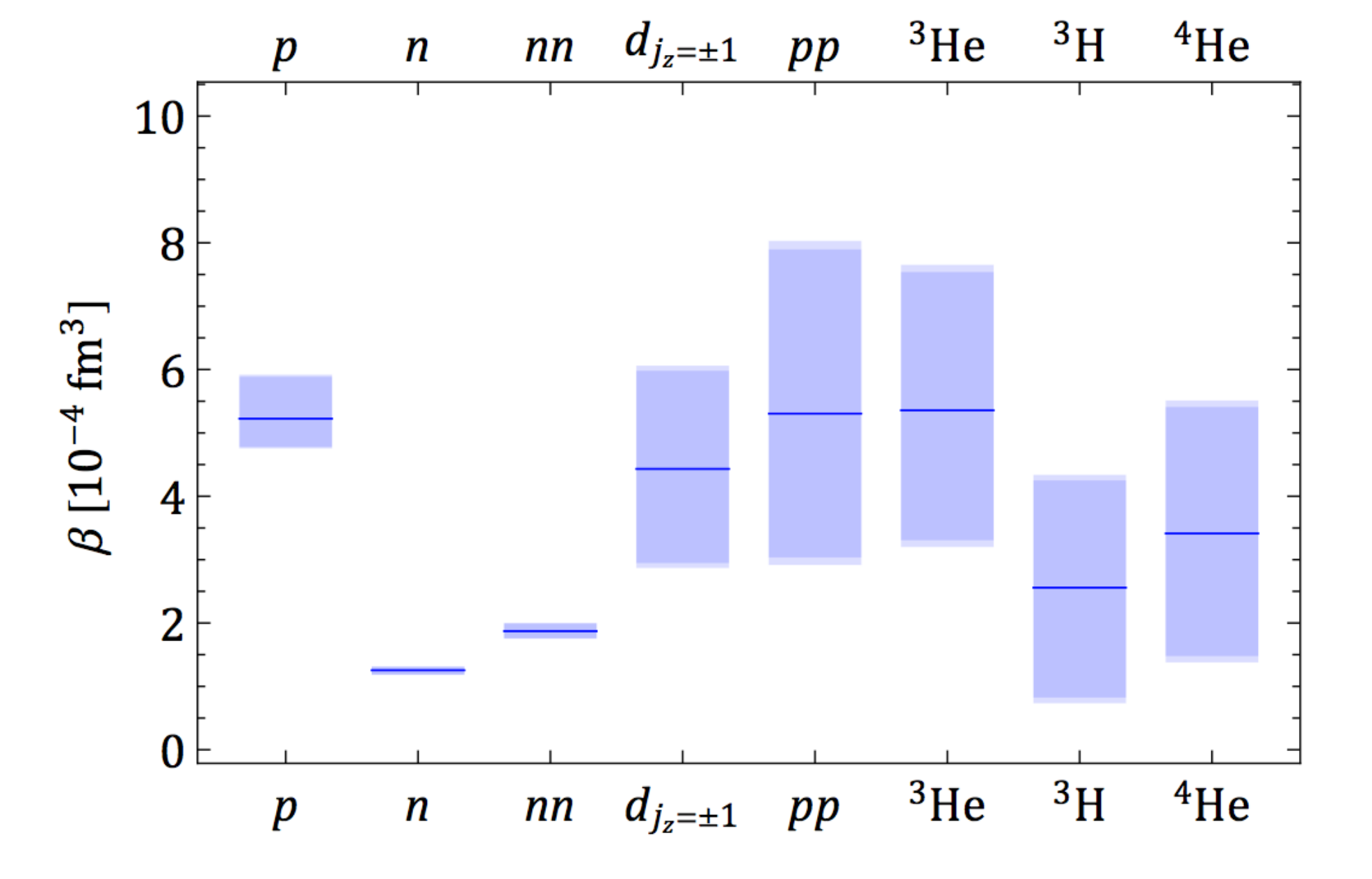}
	\caption{ 
		A summary of the magnetic moments~\protect\cite{Beane:2014ora} (left panel) and polarizabilities (right panel) of the nucleons and
		light nuclei calculated with LQCD at a pion mass of
		$m_\pi\sim 805~{\rm MeV}$~\protect\cite{Chang:2015qxa}.    }
	\label{fig:summaryBETA}
\end{figure}

\begin{figure}
	\vspace*{-0cm}
	\centering
	\includegraphics[width=0.4\columnwidth]{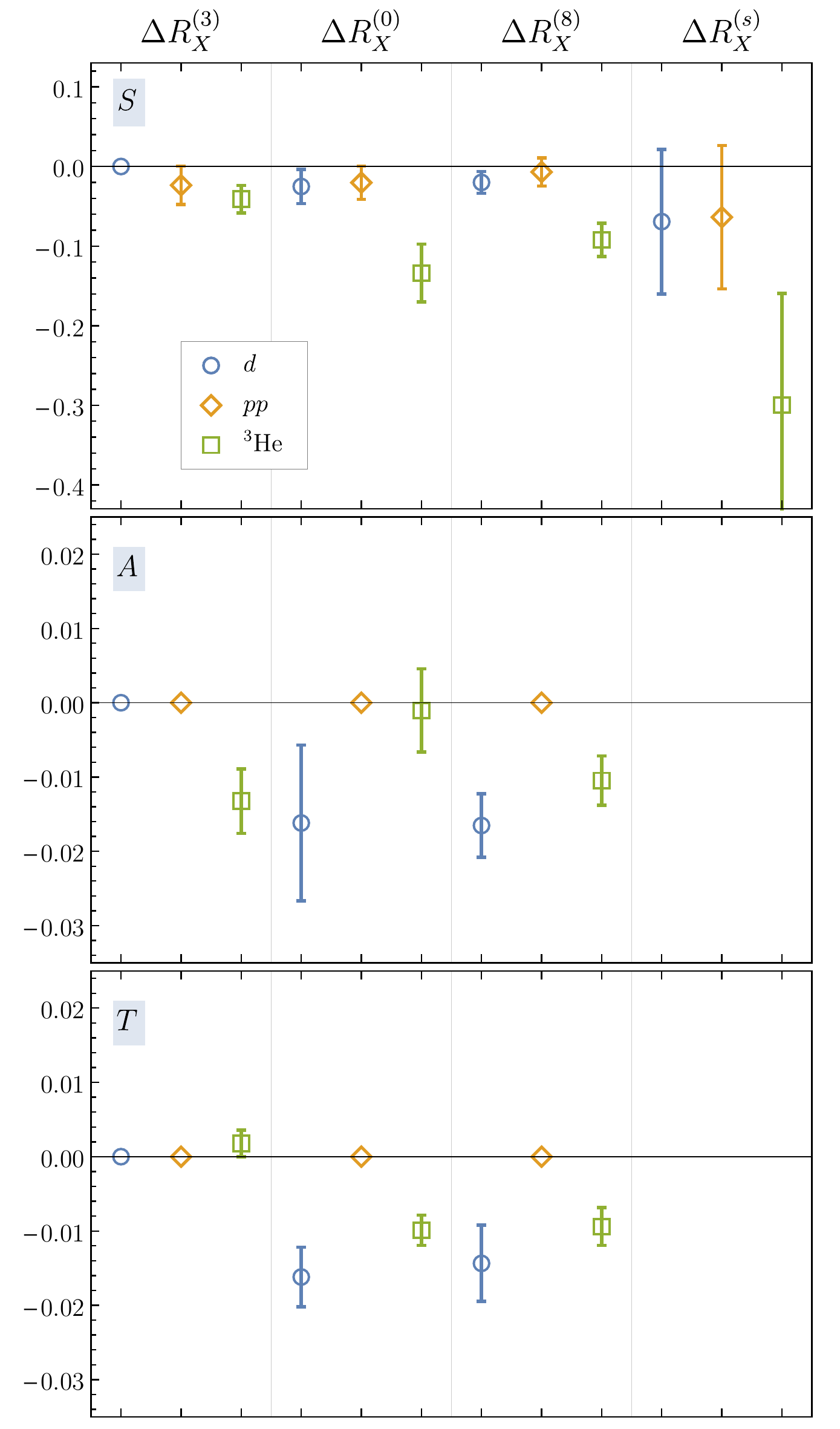}  
	\caption{ 
		The differences of the scalar, axial and tensor current matrix elements of light nuclei from single particle expectations ~\protect\cite{Chang:2017eiq}. Nuclear effects can be identified by the deviation of the quantities shown from zero.   
	}
	\label{fig:SAT}
	\vspace*{-0.5cm}
\end{figure}
The first steps towards understanding nuclear structure form LQCD have been taken, with isovector magnetic moments \cite{Beane:2014ora,Beane:2015yha,Detmold:2015daa} and magnetic  polarizabilities \cite{Chang:2015qxa} of nuclei up to $A=4$ being computed at heavier than physical quark masses using background field methods. Interestingly, relations that exist between magnetic moments in phenomenology are also apparent in the LQCD results at heavy quark masses. For example,  the magnetic moment of the triton is very close to the magnetic moment of the neutron; this is in line with the simplest shell-model configuration where the two protons in the triton spin-pair to zero. The extracted magnetic moments and polarizabilities are summarized in Fig.~\ref{fig:summaryBETA} and, as seen in left panel, the close agreement between the LQCD  and experimental magnetic moments is striking.

The Gamov-Teller (GT) contributions to the weak decay of the triton \cite{Savage:2016kon} and the coupling on $A\le3$ nuclei to scalar and tensor currents \cite{Chang:2017eiq} have also been investigated using similar background field techniques. The weak decay of the triton is the simplest nuclear probe of weak interactions and the GT matrix element is most uncertain contribution. 
As discussed extensively in the companion White Paper on Fundamental Symmetries \cite{wpfund}, the scalar current is relevant for interactions with nuclei in many models of dark matter~\cite{Undagoitia:2015gya} and for searches for new physics in precision spectroscopy~\cite{Delaunay:2016brc,Delaunay:2017dku}, while the tensor current determines 
the quark electric dipole moment contribution to a nuclear electric dipole moment~\cite{Engel:2013lsa,Yamanaka:2016umw,Chupp:2017rkp} and is thus an important ingredient in searches for new sources of time-reversal violation. 
 The nuclear dependencies of the various currents are shown in Fig.~\ref{fig:SAT}.

While the quark structure of nucleons and nuclei is relatively well probed by electron scattering experiments, unraveling the gluon structure is  more difficult. As discussed in Section \ref{sec:hadronstructure}, the Electron Ion Collider \cite{Accardi:2012qut}, a major new Nuclear Physics accelerator facility planned for construction in the 2020s, will particularly target the gluon structure of nucleons and nuclei. LQCD calculations can play an important role in planning this facility and setting benchmarks for first measurements of various gluon structure quantities. To this end, a preliminary study of the modification of the lowest moment of the unpolarized  gluon distributions in nuclei, the gluon momentum fraction, has been performed \cite{Winter:2017bfs}, although nuclear effects were bounded rather than resolved.   In addition, the first moment of the gluon transversity structure function was investigated in the spin-1 deuteron. This latter  quantity  corresponds to a target helicity flip by two units and so vanishes for the nucleon; it is therefore intrinsically a nuclear effect.

Calculations of electroweak interactions with nuclei that include momentum transfer from the currents will  determine the  nuclear form factors necessary to constrain  elastic electron-nucleus and  neutrino-nucleus scattering. This will reduce the theoretical uncertainties inherent in the coming long-baseline neutrino experimental program as discussed in detail in the companion White Paper on Neutrino Interactions \cite{wpneutrino}. Coupled to the calculations of the proton charge radius described above, calculations of the charge radii of the light nuclei $d$, $^3$He and $^4$He  will enable further insight into the discrepancies in nuclear charge radii between muon spectroscopy and electron scattering and spectroscopy \cite{Hill:2017wzi}.

Future calculations will explore the modification of moments of parton distributions in light nuclei, thereby probing the EMC effect from QCD. While these calculations will be in light nuclei for the foreseeable future, effective field theory \cite{Chen:2004zx,Chen:2016bde} and phenomenology \cite{Hen:2016kwk} suggest that two-body correlations that can be determined in the few nucleon sector are sufficient to describe the EMC effect. LQCD can also help address more complex questions such as the flavor and spin dependence of the EMC effect that are hard to access from phenomenology. Using the techniques described in Section \ref{sec:hadronstructure}, the Bjorken-$x$ dependence of nuclear parton distributions will also be accessible, significantly expanding the connection of LQCD to phenomenology in this area.

\subsection{Nuclear interactions}

Understanding the  forces between nucleons that result in their binding into nuclei is a central goal of nuclear physics. As discussed in Section \ref{sec:hadronspectroscopy} above, two-particle interactions can be addressed using the finite volume formalism of L\"uscher \cite{Luscher:1986pf,Luscher:1990ck} that translates finite volume energy levels into determinations of scattering phase shifts up to inelastic thresholds. Over the last decade, calculations of scattering phase shifts for baryon-baryon systems have become increasingly advanced, although they are still far from the level of sophistication that has been achieved in the meson sector. The two different nucleon-nucleon spin channels have been studied over a range of quark masses, with the most recent calculations  performed at quark masses corresponding to $m_\pi\sim 400$ MeV. Hyperon-nucleon scattering parameters have also been extracted and extrapolated to the physical quark masses.  
Knowledge of the nucleon-hyperon ($n\Lambda$ and $n\Sigma$) scattering phase shifts is  important in determining the equation of state of neutron stars, as strongly attractive interactions make it feasible for the dense core of neutron stars to relieve degeneracy pressure through hyperon production \cite{Lattimer:2000nx}. However the unstable nature of hyperons makes it very difficult to extract these phase shifts from experiment. Unlike in the $NN$ case, the scattering phase shifts extracted from LQCD rival the precision of phenomenological determinations and indicate that hyperons are potentially relevant in the interior of neutron stars \cite{Beane:2012ey,Wagman:2017tmp}. With recent detection of gravitational wave signatures of a  neutron star merger event by advanced LIGO and the associated electromagnetic follow-up observations \cite{TheLIGOScientific:2017qsa,GBM:2017lvd}, and with the first release of observations from the NICER satellite observatory expected soon \cite{Ozel:2015ykl}, QCD input into the nuclear equation of state has taken on a new impetus. In the coming decade such scattering phase shifts will be extracted with full control of systematic uncertainties; calculations of the nucleon-nucleon interaction will benchmark LQCD, while those in more exotic channels will be predictions that challenge experiment and act as input to phenomenology. Future calculations will also include the effects of QED in scattering analyses \cite{Beane:2014qha}.

Three-nucleon forces can also be determined from finite volume spectroscopy.   The complexity of the three-body interactions, however, means that the amplitude-based methods used for two-particle systems are challenging to apply. While the simplest aspects of the formalism needed to extract interactions from three particle finite-volume energies have been developed \cite{Beane:2007qr,Detmold:2008gh,Tan:2007bg,Kreuzer:2010ti,Polejaeva:2012ut,Briceno:2012rv,Hansen:2014eka}, as yet the only systems that have been analysed numerically are multi-meson system that interact weakly \cite{Beane:2007es}. At present, analysis of three baryon systems must resort to  effective field theory based methods \cite{Detmold:2012wpa,Detmold:2013wda,Detmold:2015jda,Kirscher:2015tka,Davoudi:2017ddj,Savage:2016egr} in which the finite volume eigenspectrum of QCD in the relevant quantum number systems is matched to EFT calculations in the same volume, thereby enabling extraction of the relevant low energy constants. This approach makes use of the full statistical power of the LQCD calculations, but relies on the EFT to extract infinite volume physics. Initial EFT studies matching to multi-nucleon ground state binding energies have been undertaken \cite{Barnea:2013uqa,Bansal:2017pwn,Contessi:2017rww,Gandolfi:2017arm}, determining the LECs of the two and three-nucleon interactions in pionless (at a set of unphysical quark masses) and pionful EFTs. Having determined these LECs, the EFTs have been used to extrapolate to larger systems such as $^{16}$O. LQCD calculations on magnetic properties have also been used to constrain EFT approaches \cite{Kirscher:2017fqc}.

The electroweak interactions of two nucleon systems are particularly important phenomenologically. 
\begin{figure}
	\centering
	\includegraphics[width=0.48\columnwidth]{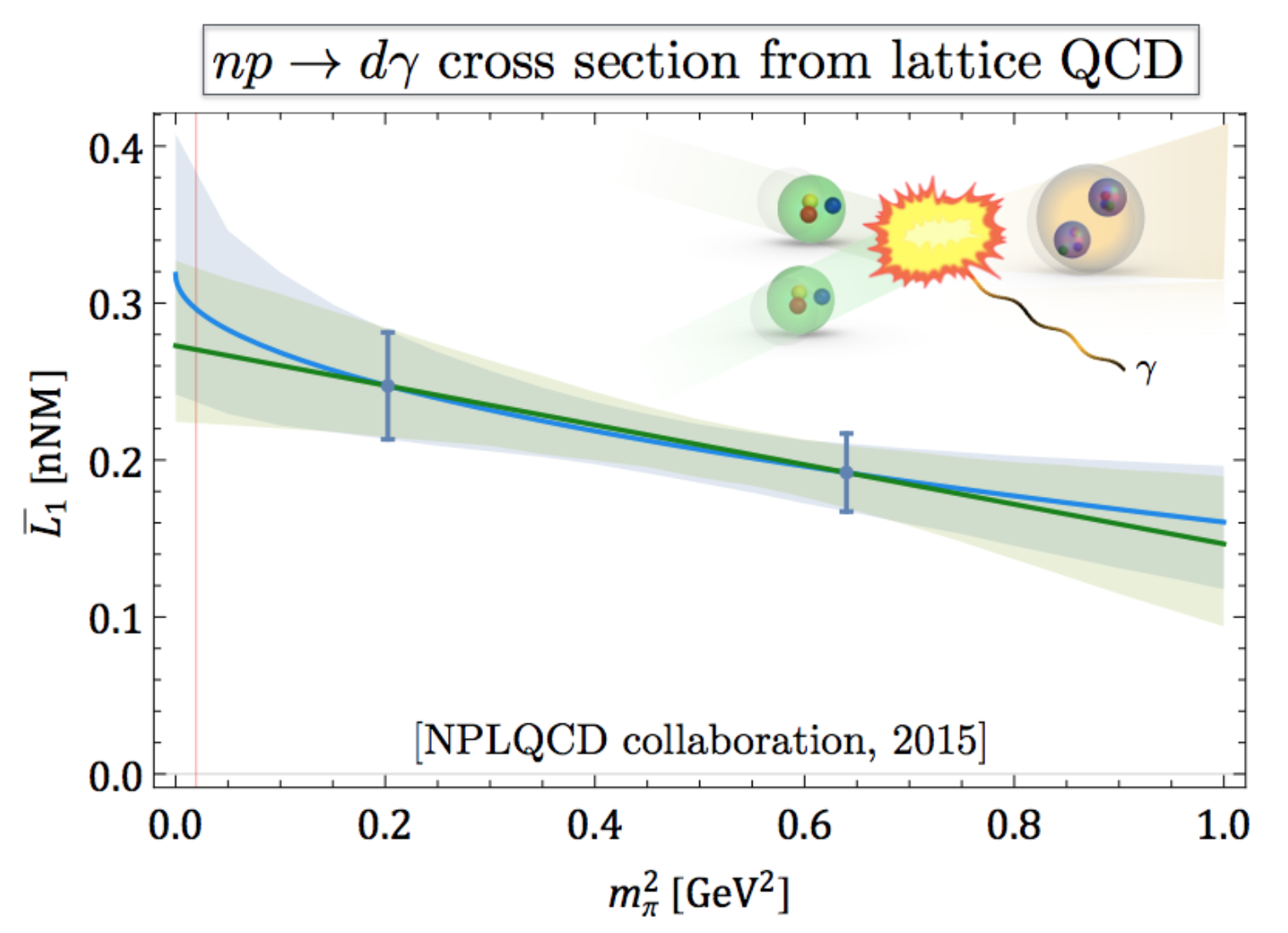}  
	\caption{ 
		The short-distance correlated two-nucleon (meson-exchange current)
		contribution to $np\rightarrow d\gamma$~\protect\cite{Beane:2015yha}.    
	}
	\label{fig:L1bar}
	\vspace*{-0.4cm}
\end{figure}
Calculations of two-nucleon systems in external magnetic fields were used to isolate the
short-distance two-body electromagnetic contributions to the low-energy radiative capture process $np\rightarrow d\gamma$,
and the photo-disintegration processes $\gamma^{(*)}d\rightarrow np$~\cite{Beane:2015yha},
as shown in Fig.~\ref{fig:L1bar}~\cite{Beane:2015yha,Detmold:2015daa}.  
In nuclear potential models, such contributions are described by 
phenomenological meson-exchange currents; using LQCD these were determined directly from the quark and gluon interactions of QCD.
This was achieved by calculations of coupled neutron-proton systems in multiple background magnetic fields, at two values of the 
quark masses, corresponding to pion masses of $m_\pi\sim 450$ and 806 MeV. The results were extrapolated to the physical pion mass, allowing the rate of the low-energy inelastic process to be determined 
at the physical point. 
This is the first LQCD calculation of an inelastic nuclear reaction.

The first LQCD calculations of the nuclear matrix element determining the $pp\rightarrow d e^+\nu_e$ fusion cross section and the Gamow-Teller matrix element contributing to tritium $\beta$ decay were presented in Ref.~\cite{Savage:2016kon}.
Using a new implementation of the background field method,
the matrix elements were calculated at the SU(3)-flavor symmetric 
value of the quark masses, corresponding to a pion mass of $m_\pi\sim 806~{\rm MeV}$. Assuming that the short-distance correlated two-nucleon contributions to the matrix element 
(meson-exchange currents) depend only mildly on the quark masses, as seen for the analogous magnetic interactions, 
the calculated $pp\rightarrow d e^+\nu_e$ transition matrix element leads to a fusion cross section at the physical quark 
masses that is consistent with its currently accepted value, although further calculations are required to  better substantiate this conclusion. Moreover, the leading two-nucleon axial counterterm of pionless EFT is determined to 
be $L_{1,A} = 3.9(0.2)(1.0)(0.4)(0.9)~{\rm fm}^3$ at a renormalization scale set by the physical pion mass, 
also in agreement with the accepted phenomenological range.

For some specific nuclei, single $\beta$ decay is energetically forbidden, but double $\beta$ decay is allowed. In the Standard Model, this decay occurs with the release of two electrons and two anti-neutrinos, conserving lepton number  ($2\nu\beta\beta$-decay). In many beyond-the-Standard-Model scenarios, either with light Majorana neutrinos (particles that are their own antiparticles) or with other forms of lepton number non-conservation at high scales,  a second form of  double $\beta$ decay that involves no neutrinos in the final state ($0\nu\beta\beta$-decay) can occur. Observation of this latter process would be an unambiguous signal for new physics. Understanding of the implications of such an observation, as well as optimizing the design of future experiments seeking this decay mode, requires understanding the relevant $\Delta I=2$ nuclear transition matrix elements. This is a challenging task and state-of-the-art nuclear theory calculations of these matrix elements differ by an order of magnitude. LQCD offers the prospect of QCD input into  this problem through calculations of the relevant matrix elements in light nuclei that can be used to control uncertainties in nuclear models. In the last two years, the $2\nu\beta\beta$ process has been studied in the $pp\to nn$ transition \cite{Tiburzi:2017iux,Shanahan:2017bgi}, the pionic matrix elements, $\langle \pi^+ | {\cal O} | \pi^- \rangle$, of $\Delta I =2$ short distance operators \cite{Nicholson:2018mwc}, and the $\pi^-\to \pi^+ e^- e^-$ and $\pi^-\pi^-\to e^-e^-$ transitions induced  by a light Majorana neutrino \cite{Feng:2018pdq,Detmold:2018zan}, have all been investigated for the first time. Future refinements of these calculations, even restricted to few nucleon systems, have the potential to significantly impact experimental design.  Already, these finding suggests that nuclear models for neutrinoful and neutrinoless $\beta\beta$ decays need to incorporate a previously neglected contribution if they are to provide reliable guidance for next-generation neutrinoless $\beta\beta$-decay searches.

\subsection{Nuclear input for neutrino physics and fundamental symmetries}

Nuclei are used as targets in intensity frontier experiments probing the neutrino sector and searching for physics beyond the SM. In particular, argon ($Z=18$) is the target material for a number of current neutrino experiments and will be the target for the upcoming Deep Underground Neutrino Experiment (DUNE), while a range of nuclei such as sodium ($Z=11$) and  xenon ($Z=54$) are used in dark matter direct detection experiments \cite{Undagoitia:2015gya}. Charge lepton flavor violation searches look for $\mu\to e$ conversion in the field of aluminium ($Z=13$) \cite{Albrecht:2013wet}, and precision isotope-shift spectroscopy experiments consider a wide range of nuclei ranging from hydrogen ($Z=1$) to ytterbium ($Z=70$) in  order to constrain new physics \cite{Delaunay:2016brc,Delaunay:2017dku}, both requiring knowledge of various nuclear matrix elements \cite{Chang:2017eiq}. Finally, double-$\beta$ decay (DBD) searches utilize heavy isotopes to search for lepton number violation through neutrinoless DBD \cite{DellOro:2016tmg,Engel:2016xgb} as discussed above.

All of the techniques discussed above in the study of nuclear spectroscopy, structure and interactions are applicable in these areas, coupled to EFT methods to reach the experimentally relevant nuclei. We leave a full discussion of these topics to the two companion USQCD white papers on Neutrino-Nucleus Interactions \cite{wpneutrino}and on Fundamental Symmetries \cite{wpfund}.

\section{Future Opportunities}
\label{sec:future}

\subsection{Hadron Structure}

Understanding the structure of the proton and other hadrons is an important facet of  nuclear science and  has led to revolutionary discoveries over the last 70 years, including that of QCD itself.  
The discrepancy in the experimental measurements of the charge radius of the proton~\cite{Antognini:1900ns} has spurred a flurry of new measurements and phenomenological estimates. Current lattice QCD determinations of the proton radius are challenging since the derivative which defines the radius is extracted from modeling the $Q^2$ dependence of form factors from calculations at the discrete values of momentum accessible in a finite lattice volume.  Indeed, the problems encountered mirror those in electron-scattering experiments, where the form factor is computed for a discrete, albeit closely spaced, set of finite $Q^2$, and the need to include dispersive methods in the analysis of the form factors over the values of $Q^2$ probed in experiment has been emphasised~\cite{Alarcon:2018irp}.

A method to avoid these uncertainties involves the computation of coordinate-weighted moments of currents, without the need to model the $Q^2$ dependence~\cite{Bouchard:2016gmc}. Another recent approach~\cite{Detmold:2018ptb} introduces a mass splitting between the up and down quarks, allowing to access time-like as well as space-like four-momentum transfers close to $Q^2=0$. These calculations with statistical precision on the order of a few percent, and at the physical quark mass and several lattice spacings, are conceptually straightforward and achievable in the near term. The inclusion of disconnected diagrams would allow the access to the proton and neutron charge radii directly.

The nucleon electric and magnetic form factors $G_{E,M}(Q^2)$ describe the distribution of charge and
magnetization inside the
nucleon~\cite{Burkardt:2000za,Burkardt:2002hr,Miller:2007uy,Carlson:2007xd} as a function of the momentum, $Q$, carried by the photon probe.
They have been extensively studied since the dawn of
accelerator technology, offering the first experimental evidence for composite structure 
of nucleons~\cite{Hofstadter:1955ae}, as well as the first
determination of the proton radius~\cite{Chambers:1956zz},
and remain an area of active experimental and theoretical research. The high-momentum limit yields a ``high-resolution'' picture of the nucleon, and is a subject of experiments at JLab.

Calculations involving hadrons with large momentum introduce unique challenges in LQCD: in the Breit frame one 
has to study nucleon states with high momentum 
resulting in statistical noise as well as excited state contributions due to the
shrinking energy gap between the nucleon states.
New techniques, such as momentum smearing~\cite{Bali:2016lva,Syritsyn:2017jrc}, have been shown to improve
the signal for boosted nucleon correlators by a factor of at least 10. The efficacy of the approach suggests precision calculations of nucleon form-factors are achievable, up to a few GeV$^2$ in a few years.

The recent development of techniques allowing for the extraction of the Bjorken-$x$ dependence of quark and gluon distribution functions directly from Euclidean space calculations~\cite{Ji:2001wha} has opened the door to a new age of hadronic and nuclear physics calculations. In particular, these methods allow, in principle,  the extraction of the large-$x$ dependence of quark and glue distributions that are a subject of the 12 GeV upgrade at JLab, and the small-$x$ dependence under study at RHIC experiment at BNL and the proposed EIC facility. Near-term  lattice calculations will establish the techniques and quantify systematic uncertainties for LQCD studies of the $x$-dependence of PDFs within systems like the pion, kaon and the nucleon. Establishing the flavor dependence of such parton distribution functions is also a near term goal. However, accessing the small-$x$ dependence, where the glue is expected to dominate, is a challenging goal, as naively one expects that very large lattice sizes with small lattice spacings are required. An intermediate approach could use anisotropic lattices to decrease the lattice spacing in a spatial direction.

\subsection{Hadron Spectroscopy}

One goal of the spectroscopy program is establishing the branching fractions for decays of hadrons, including putative exotic mesons. It is these decay couplings that can inform and confront experimental analyses, such as those within GlueX and CLAS12. A target within the next few years is establishing the spectrum of the low-lying scalar, vector and tensor resonances in the physical limit of QCD. While these calculations must be mindful of potential three body decays, they are achievable within the next few years using resources available on leadership computing resources as well as USQCD resources.

Targeting exotic meson decays is more challenging, particularly because three-body decays might well be important. First calculations will necessarily need to use unphysically large quark masses where three-body thresholds are pushed higher in the spectrum and away from the resonance region of interest.
These initial calculations are tractable in the near term. However, the inclusion of three-body decays within searches for exotics is more challenging. The computational cost can be addressed with anticipated improvements of algorithms, but the understanding of how three- and higher-body decay channels can be included is more a conceptual question that needs to be addressed.

Electromagnetic radiative transitions provide a probe into the structure of resonant states, and while challenging, are experimentally accessible. 
One notable target is the photo excitation of exotic mesons from pion exchange off the proton, the experimental production mechanism for the GlueX experiment. Thus, the extraction of the photo-production rate of exotic mesons is an important target for lattice calculations as they can inform the analysis of on-going experiments. The analytical formalism for the study of composite states exists~\cite{Briceno:2015tza}, and first studies have been carried out; however, the analytic formalism is not in place for systems with three body decays. To avoid such complications, first calculations will proceed at unphysical pion masses;  these studies are achievable in the near term. A successful extraction of a photo-coupling will be an important step for phenomenology.

Beyond reproducing experimentally accessible reactions, lattice calculations can also investigate physically relevant quantities that cannot be determined experimentally. First such calculations will include the elastic form factors of hadronic resonances. Encoded in these is structural information, which will give further insight into the true nature of these states, e.g., their size and shape. The computational aspects are manageable, while the analytic formalism for such systems is maturing~\cite{Briceno:2015tza,Baroni:2018iau}.

A compelling question that remains unanswered in the charmonium sector is the nature of the `XYZ' resonances. They often appear in close proximity to thresholds leading to wide-ranging speculations that some of these might be `molecular' in origin. These questions can be tested by studying the response of the state to variations of the position of the threshold induced by changing the light and charm quark masses. In addition, the behavior of radiative transition decays and form-factors, including the calculation of charge radii, will provide valuable insight into their nature. Necessarily, the calculations will involve a range of light and charm quark masses, both at and away from their physical values..  These computations are relatively straightforward and achievable within the next few years. However, some of the systems, such as the $X(3872)$, are very close to threshold thus potentially requiring the inclusion of isospin breaking effects as well as high statistical precision.

\subsection{Nuclear Interactions}

Since nuclei make up the majority of the visible matter in the Universe, understanding their emergence from the underlying theory of the strong interaction is a fundamental challenge bridging nuclear and particle physics. Large-scale numerical calculations will allow us to address this challenge and achieve a quantitative connection between the Standard Model and nuclear phenomenology, opening new directions in our quest to interpret the complexities of nuclear physics and supporting experimental efforts to use nuclei to reveal fundamental aspects of nature.

Determinations of the finite-volume energy levels of few nucleon systems constrain the two- and three- nucleon forces in EFT methods~\cite{Barnea:2013uqa}, thereby enabling predictions of the properties and interactions of larger nuclei. Calculations of the spectrum of light nuclei with atomic number $A\le 4$ with quark masses close to the physical limit are achievable in the near term, given ${\cal O}(10^5)$ quark propagator sources, coupled with techniques such as the variational method~\cite{Michael:1985ne}, improved estimators \cite{Beane:2014oea}, along with signal-to-noise optimization methods~\cite{Detmold:2014hla}, all to enhance the statistical signal. While not fully resolving all systematic uncertainties, however, these calculations are expected to represent a significant step forward in showing how nuclei emerge from the intricacies of Standard Model dynamics.

The extension to the larger $p$-shell nuclei will be important in the future as they are more sensitive to three body nuclear forces than lighter nuclei. They also present a new level of challenge for calculations as their structure is more complicated. These systems will require significantly more sophisticated constructions of interpolating operators such as determinant contraction methods~\cite{Detmold:2012eu,Vachaspati:2014bda}. A large number of quark propagators will be required and the resulting cost of the contractions of these propagators will be large as well. Current development efforts are underway that will allow  first tests of the efficacy of the approach, initially at unphysical pion masses, in the near term.

As outlined in this and accompanying white-papers, there are strong phenomenological motivations for studies of the scalar, axial and tensor current matrix elements of light nuclei. Calculations of the matrix elements in light nuclei up to $A=4$ \cite{Winter:2017bfs} at close-to-physical values of the quark masses, will constrain necessary inputs for current and future experiments using nuclear targets for searches for physics beyond the Standard Model. The statistical requirements for each calculation will depend on the magnitude of the signal in each channel, which is currently unknown. First studies are expected to proceed with the scalar current as it appears to have the strongest nuclear effects. Connected contributions can use external field techniques~\cite{Savage:2016kon,Shanahan:2017bgi,Tiburzi:2017iux} while disconnected contributions are expected to be important. These calculations will establish the baseline for the statistics required, and while computationally challenging, are achievable in a few years time.

\begin{figure}
    \vspace*{3cm}
\end{figure}

\bibliographystyle{apsrev4-1}
\bibliography{paper}

\end{document}